\documentclass[aps,prl,twocolumn,showpacs,floatfix,superscriptaddress, longbibliography]{revtex4-2}
\usepackage[]{graphicx}
\usepackage{epstopdf}
\usepackage{ulem} % Add this in the preamble
\usepackage{float}

\usepackage{graphicx}% Include figure files
\usepackage{dcolumn}% Align table columns on decimal point
\usepackage{bm}% bold math
\usepackage{graphicx}
\usepackage{xcolor}
\usepackage{amsmath}
\usepackage{amssymb}
\usepackage{siunitx}
\usepackage{lipsum} 
\usepackage{enumerate}
\usepackage[colorlinks=true]{hyperref}
\usepackage[dvipsnames]{xcolor}

\usepackage{lineno}

\usepackage{soul}

\begin{document}

\title{Efficient photon-pair emission from a nanostructured resonator and its theoretical description}

\author{Michael Poloczek}
\altaffiliation{These authors contributed equally to this work.}
\affiliation{Friedrich-Alexander-Universität Erlangen-Nürnberg, Staudtstr. 7, 91058 Erlangen, Germany}
\affiliation{Max Planck Institute for the Science of Light, Staudtstr. 2, 91058 Erlangen, Germany}

\author{Alberto Paniate}
\altaffiliation{These authors contributed equally to this work.}
\affiliation{Friedrich-Alexander-Universität Erlangen-Nürnberg, Staudtstr. 7, 91058 Erlangen, Germany}
\affiliation{Max Planck Institute for the Science of Light, Staudtstr. 2, 91058 Erlangen, Germany}
\affiliation{Quantum metrology and nano technologies division, INRiM,  Strada delle Cacce 91, 10135 Torino, Italy}

\author{Attilio Zilli}
\affiliation{Department of Physics, Politecnico di Milano, 20133 Milan, Italy}

\author{Vitaliy Sultanov}
\affiliation{Friedrich-Alexander-Universität Erlangen-Nürnberg, Staudtstr. 7, 91058 Erlangen, Germany}
\affiliation{Max Planck Institute for the Science of Light, Staudtstr. 2, 91058 Erlangen, Germany}

\author{Yigong Luan}
\affiliation{Department of Physics, Politecnico di Milano, 20133 Milan, Italy}

\author{Tomás Santiago-Cruz}
\affiliation{Friedrich-Alexander-Universität Erlangen-Nürnberg, Staudtstr. 7, 91058 Erlangen, Germany}
\affiliation{Max Planck Institute for the Science of Light, Staudtstr. 2, 91058 Erlangen, Germany}

\author{Luca Carletti}
\affiliation{Department of Information Engineering, University of Brescia, Via Branze 38, 25123 Brescia, Italy}

\author{Marco Finazzi}
\affiliation{Department of Physics, Politecnico di Milano, 20133 Milan, Italy}

\author{Marco Genovese}
\affiliation{Quantum metrology and nano technologies division,  INRiM,  Strada delle Cacce 91, 10135 Torino, Italy}

\author{Ivano Ruo-Berchera}
\affiliation{Quantum metrology and nano technologies division,  INRiM,  Strada delle Cacce 91, 10135 Torino, Italy}

\author{Marzia Ferrera}
\affiliation{Istituto Italiano di Tecnologia, Via Morego 30, Genova 16163, Italy}

\author{Andrea Toma}
\altaffiliation{Present address: \textit{Dipartimento di Fisica, Università di Genova, Via Dodecaneso 33, 16146 Genova, Italy.}}
\affiliation{Istituto Italiano di Tecnologia, Via Morego 30, Genova 16163, Italy}

\author{Francesco Monticone}
\affiliation{School of Electrical and Computer Engineering, Cornell University,
Ithaca, New York, 14850, USA}

\author{Michele Celebrano}
\email{michele.celebrano@polimi.it}
\affiliation{Department of Physics, Politecnico di Milano, 20133 Milan, Italy}

\author{Maria Chekhova}
%\email{maria.chekhova@mpl.mpg.de}
\affiliation{Friedrich-Alexander-Universität Erlangen-Nürnberg, Staudtstr. 7, 91058 Erlangen, Germany}
\affiliation{Max Planck Institute for the Science of Light, Staudtstr. 2, 91058 Erlangen, Germany}

\begin{abstract} 	
Spontaneous parametric down-conversion (SPDC) in subwavelength nanostructures is a promising source of quantum light, owing to its multifunctionality and ability to generate complex quantum states. Nevertheless, the mechanisms governing photon-pair generation in such systems remain only partially understood. In particular, experimental investigations of key emission properties in individual resonators, such as directionality and spectral distribution, are still lacking, and predictive theoretical frameworks have not yet been experimentally validated. Here, we report the first measurement of the directional and spectral distributions of photon-pairs generated via SPDC in a nanostructured resonator. Both distributions exhibit resonant behaviour, which we describe using an extended quasi-normal-mode theory. This comparison is enabled by photon-pair count rates of up to 0.45 Hz/mW-- to our knowledge, the highest reported for a nanostructured resonator. Our results provide new physical insight into nanoscale SPDC and represent an important step toward designing efficient miniaturized quantum light sources.
\end{abstract}	

\maketitle
\section{Introduction}
Traditionally, spontaneous parametric down-conversion (SPDC) is exploited in bulk non-centrosymmetric crystals to generate pairs of entangled photons, commonly referred to as signal (s) and idler (i), with frequencies satisfying energy conservation. Few millimetre-sized crystals yield high efficiencies of SPDC as they provide a large interaction length. However, in recent years, SPDC at the nanoscale has attracted significant interest as a route towards compact and integrable sources of quantum light \cite{ding2023advances,schulz2024roadmap,Ma2024}. Beyond miniaturization, the relaxation of phase-matching constraints in nanoscale systems enables intrinsic multifunctionality, in particular, the generation of highly versatile quantum states with broad spectral and angular bandwidths \cite{okoth2019microscale, okoth2020idealized}, arbitrary and tunable polarization entanglement \cite{sultanov2022flat, ma2023polarization, noh2024quantum, ma2025nonlinearity,liang2025, jia2025polarization}, spatial entanglement \cite{zhang2022spatially}, bi-directional emission \cite{son2023photon, weissflog2024directionally,lu2025}, and more complex quantum-state engineering \cite{santiago2022resonant} including the observation of two-photon quantum interference \cite{noh2025fano}. These capabilities have been demonstrated across a variety of nanophotonic platforms, including resonant metasurfaces \cite{santiago2021photon, zhang2022spatially, santiago2022resonant, son2023photon, noh2024quantum, weissflog2024directionally, noh2025fano, ma2025quantum, jia2025polarization, ma2025nonlinearity}, subwavelength nonlinear films \cite{santiago2021entangled, sultanov2022flat, stich2026thin,liang2025,lu2025}, and dielectric nanoresonators \cite{marino2019spontaneous, duong2022spontaneous, saerens2023background}. Importantly, nanoscale SPDC enables the use of a wide range of highly nonlinear materials, including emerging platforms such as van der Waals crystals \cite{guo2023ultrathin, weissflog2024tunable, kallioniemi2025van, trovatello2025quasi} or organic materials like liquid crystals \cite{sultanov2024tunable}.

Despite these advances, nanostructured SPDC sources still face fundamental challenges. Most notably, there is a lack of a quantitative description of their key emission properties, including  directionality, spectral distribution, and the generation efficiency. Moreover, a comparison between theoretical predictions and experimental measurements is still missing. This has hindered systematic device optimization, resulting in low generation efficiencies. 

\begin{figure*}
    \centering
    \includegraphics[width=\textwidth]{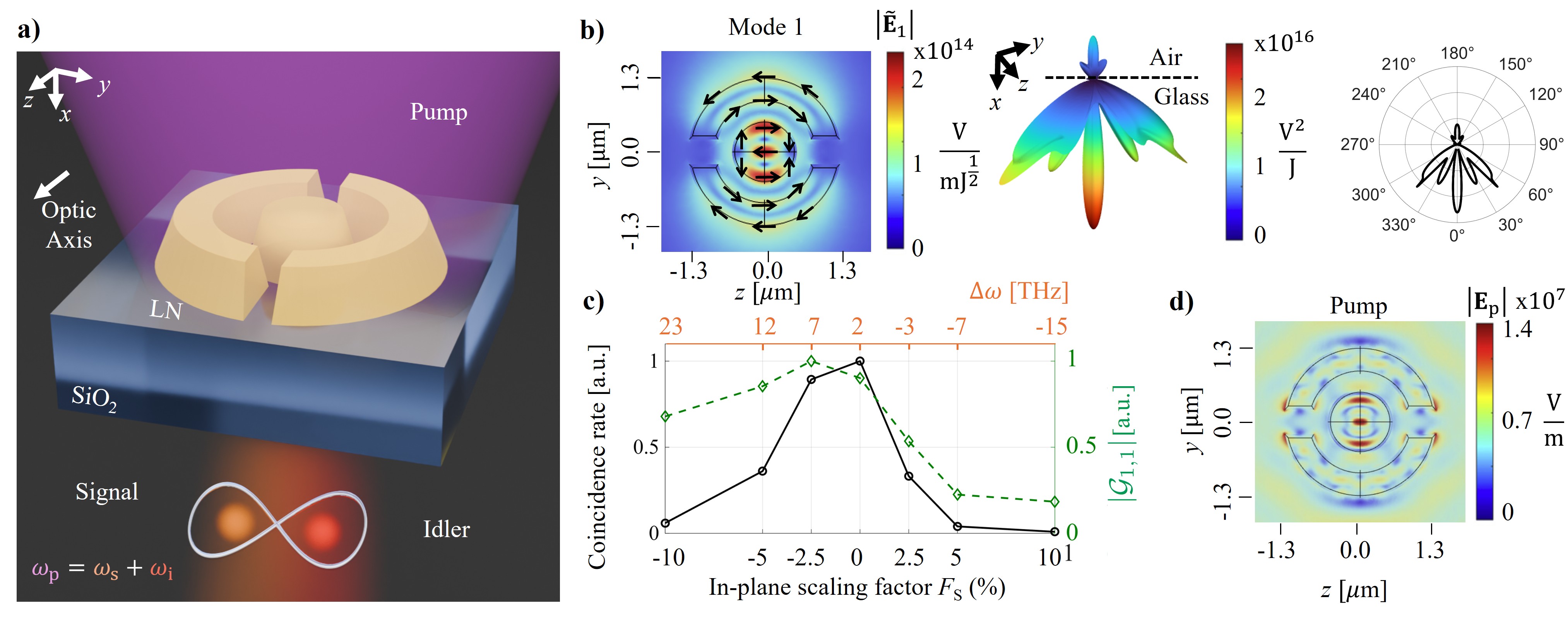}
    \caption{\textit{Spontaneous parametric down-conversion in a lithium niobate nanostructured resonator.} (\textbf{a}) Conceptual picture of the entangled photon-pair generated by the nanoresonator. (\textbf{b}) Near-field distribution (left panel) and far-field angular radiation pattern \(|\mathbf{\tilde{E}}(\theta, \varphi)|^2\) (right panel, 3D and polar visualization on the \(x-y\) plane) of the chosen quasi-normal mode \(\text{QNM}_1\). The black arrows indicate the direction of the electric field inside the resonator.  (\textbf{c}) Simulated normalized coincidence rate as a function of the in-plane scaling parameter \(F_\mathrm{S}\), i.e. a uniform percentage change of the lateral dimensions along the \(y\) and \(z\) directions with respect to the geometry used in the experiment (\(F_\mathrm{S}=0\)).  The factor \(F_\mathrm{S}\) affects the detuning \(\Delta\omega = (\omega_1 - \omega_{\mathrm{deg}})/(2\pi)\) between the eigenfrequency \(\omega_1\) of QNM\(_1\) and the degenerate angular frequency \(\omega_{\mathrm{deg}}\), as well as the spatial overlap \(\mathcal{G}_{1,1}\), shown with the green curve. (\textbf{d}) Pump electric field modulus \(|E_\mathrm{p}(\mathbf{r})|\), shown as a steady-state profile, calculated for a plane wave with the intensity \(I_0 \approx 10^{11}\,\mathrm{W/m^2}\) incident from the resonator side and polarized along the  \(z\)-axis.}
    \label{fig:concept}
\end{figure*}

In nanostructured systems, a large number of interdependent variables %play a crucial role in 
determine the properties of photon-pair emission. These include the spatial distribution of the pump electric field, the second-order susceptibility tensor \(\chi^{(2)}\),  the presence of multiple resonant modes at the signal and idler frequencies, and their mutual interference. As a result, emission properties such as directionality and spectral response cannot be adequately described within the traditional analytical frameworks of nonlinear quantum optics~\cite{klyshko2018photons, sorensen2025simple} and are typically accessible only through full-wave electromagnetic numerical simulations. In this context, most theoretical studies have relied on the correspondence between SPDC and the sum-frequency generation (SFG)~\cite{parry2021enhanced, jin2021efficient, liu2024efficient, mazzanti2022enhanced}. Although this approach is widely adopted, it provides limited physical insight into the underlying modal interactions, makes the reconstruction of the full angular and spectral emission cumbersome, and, while it has been experimentally validated in waveguides \cite{lenzini2018direct}, its validation in nanoresonators and metasurfaces remains only partial. More recently, a theoretical framework based on Green’s functions expanded in terms of quasi-normal modes (QNMs)~\cite{kristensen2014modes,lalanne2018light, gigli2020quasinormal, sauvan2022normalization} has been developed and applied to SPDC in a single nanoresonator \cite{poddubny2016generation, weissflog2024nonlinear}; however, this theory has mainly considered idealized scenarios, neglecting effects such as the presence of a substrate or collection into a single-mode fiber (SMF). Moreover, it has not yet been benchmarked against experimental results.

In this work, we design and fabricate a nanostructured resonator whose high efficiency for SPDC allows us to perform, for the first time, the directionally and spectrally resolved measurements of photon-pair emission in different scenarios. Furthermore, we extend the QNM model to account for the substrate and the coupling into an SMF, enabling a direct comparison between theoretical predictions and experimental measurements and providing a partial validation of the model. 

\section{Design and simulation}

Our device is inspired by a circular Bragg grating, also known as bullseye nanostructured resonator, which in previous works~\cite{davanco2011circular, liu2019solid, wang2019demand, barbiero2024polarization} was fabricated upon individual quantum dots to extract and redirect their single- and two-photon emission. In contrast, our design unifies the emitter and the outcoupler in a single monolithic structure. Moreover, our resonator footprint is minimized, as it  consists of a central cylinder surrounded by a single concentric ring with two radial cuts (Fig.~\ref{fig:concept}(a)), designed initially to suppress whispering-gallery modes while supporting controlled out-of-plane outcoupling. It is fabricated of lithium niobate (LN) and has a total radius of $1.3~\mu\mathrm{m}$ and an inner-cylinder radius of 600 $\mathrm{nm}$ with a  thickness of $500~\mathrm{nm}$. We pump the nanoresonator at 725 nm from the air side and collect the generated photon-pairs through the fused silica substrate. 

\begin{figure*}
    \centering
    \includegraphics[width=\textwidth]{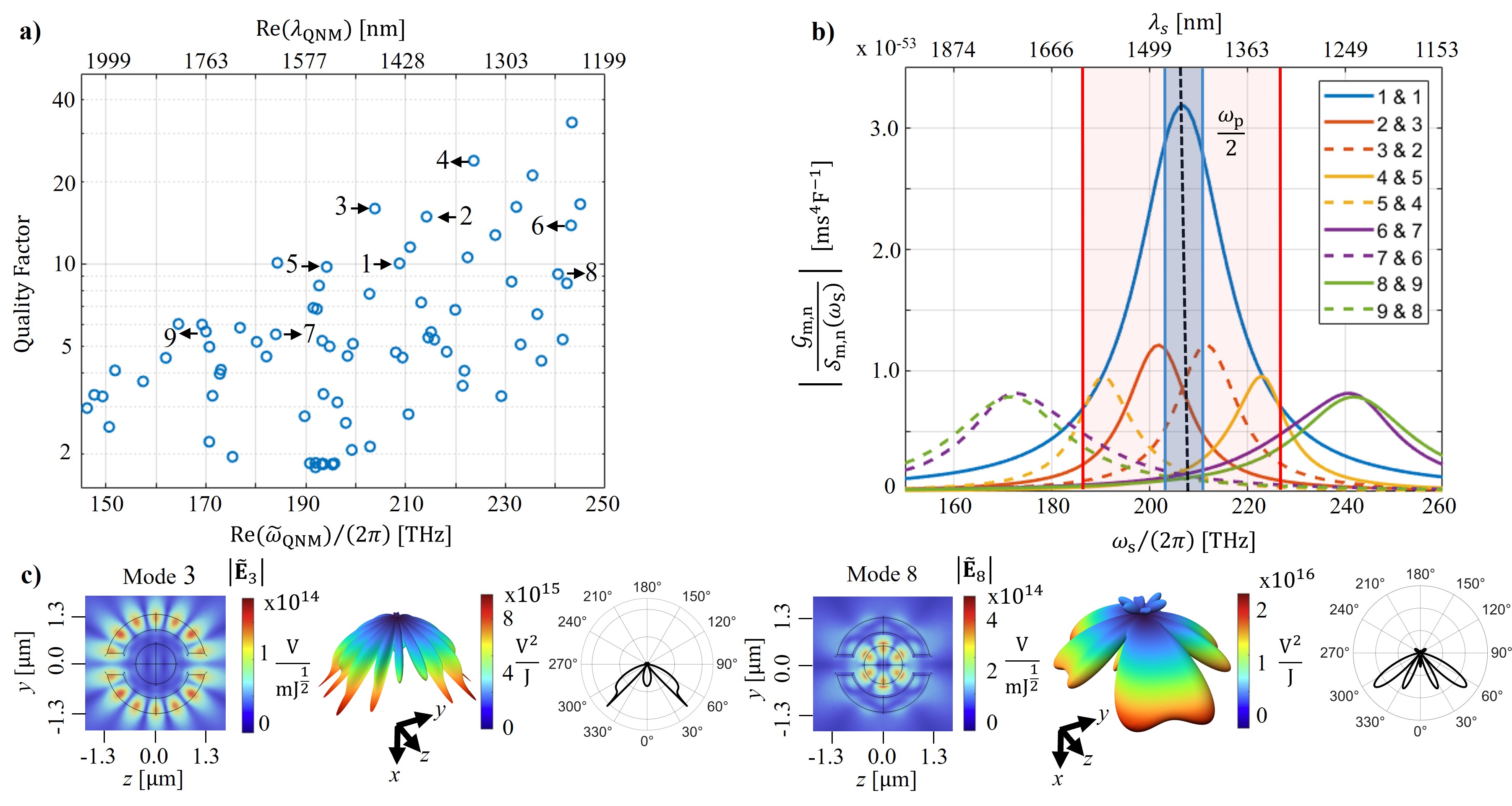}
    \caption{\textit{Quasi-normal modes of the resonator and modal-overlap coefficient. }(\textbf{a}) Quality factors of all QNMs of the nanostructure, plotted versus the real parts of their eigenfrequencies. Labels indicate the modes exhibiting the largest ratio \(|\mathcal{G}_{\mathrm{m,n}}/\mathcal{S}_{\mathrm{m,n}}(\omega_\mathrm{s})|\) across the reported range of frequencies. (\textbf{b}) Ratio \(|\mathcal{G}_{\mathrm{m,n}}/\mathcal{S}_{\mathrm{m,n}}(\omega_\mathrm{s})|\) for the nine QNM pairs with the highest peak values. %The corresponding wavelengths are indicated in the upper horizontal axis. 
    The vertical dashed line marks the degenerate frequency. The red and blue shaded regions indicate the two frequency collection ranges used in the measurements shown in Fig.~\ref{fig:ExpSim}. (\textbf{c}) Near-field and far-field distributions of two typical modes, \(3\) and \(8\), illustrating their different far-field patterns and intensities.}
    \label{fig:QNMcomplete}
\end{figure*}

To interpret the SPDC emission from the nanostructured resonator, the detected photon-pair rate is described in terms of the two-photon amplitude, \(\widetilde{T}_{\mathrm{is}}\), expressed as a coherent sum of all QNM pairs \cite{weissflog2024nonlinear}:
\begin{equation}
   \widetilde{T}_{\mathrm{is}} =
   \sum_{\mathrm{m,n} = 1}^{\infty}
   \frac{\mathcal{G}_{\mathrm{m,n}}}{\mathcal{S}_{\mathrm{m,n}}(\omega_\mathrm{s})}\,
   \tilde{E}_{\mathrm{m},\mathbf{d}_\mathrm{i}}(\theta_\mathrm{i}, \varphi_\mathrm{i})\,
   \tilde{E}_{\mathrm{n},\mathbf{d}_\mathrm{s}}(\theta_\mathrm{s}, \varphi_\mathrm{s}),
\label{eq:TPA}
\end{equation}
where $\tilde{E}_{\mathrm{m},\mathbf{d}}(\theta,\varphi)$ denotes the angular dependence of the complex electric far-field of the \(m\)-th QNM, evaluated at the spherical angles $(\theta,\varphi)$ and projected onto the detection polarization \(\mathbf{d}\). Subscripts $s$ and $i$ refer to signal and idler photons, respectively, and $\omega_\mathrm{s}$ is the signal angular frequency. \(\widetilde{T}_{\mathrm{is}}\) represents the joint probability amplitude for detecting signal and idler photons at the specified angles, frequencies, and polarizations; its explicit dependence on these variables is omitted for brevity. Each QNM pair is weighted by the spatial overlap
\begin{equation}
\mathcal{G}_{\mathrm{m,n}} =
\sum_{\alpha,\beta,\gamma}
\int d\mathbf r\,
\chi^{(2)}_{\alpha\beta\gamma}(\mathbf r)\,
\tilde E_{\mathrm{m},\alpha}(\mathbf r)\,
\tilde E_{\mathrm{n},\beta}(\mathbf r)\,
E_{\mathrm{p},\gamma}(\mathbf r),
\label{eq:G_factor}
\end{equation}
and the detuning factor
\begin{equation} 
\mathcal{S}_{\mathrm{m,n}}(\omega_\mathrm{s}) =(\omega_\mathrm{p}-\omega_\mathrm{s}-\tilde\omega_\mathrm{m})\tilde\omega_\mathrm{m} (\omega_\mathrm{s}-\tilde\omega_\mathrm{n})\tilde\omega_\mathrm{n}. 
\label{eq:overlFactor}
\end{equation}
The integral \(\mathcal{G}_{\mathrm{m,n}}\) quantifies the overlap in the nonlinear material between the nonlinear susceptibility tensor \(\chi^{(2)}\), the near-field QNM eigenfields \(\tilde{\mathbf{E}}_\mathrm{m}\) and \(\tilde{\mathbf{E}}_\mathrm{n}\) along the directions \(\alpha\) and \(\beta\), and the pump field \(\mathbf{E}_\mathrm{p}\) along the direction \(\gamma\). The spectral factor \(\mathcal{S}_{\mathrm{m,n}}(\omega_\mathrm{s})\) accounts for the complex detuning between the generated photon frequencies of the idler and signal photons \((\omega_\mathrm{i}=\omega_\mathrm{p}-\omega_\mathrm{s},\,\omega_\mathrm{s})\), where \(\omega_\mathrm{p}\) is the frequency of the pump, and the QNM eigenfrequencies \((\tilde{\omega}_\mathrm{m},\tilde{\omega}_\mathrm{n})\), thereby determining the spectral and modal selectivity of SPDC. For the simulated parameters, complete model and the QNM normalization used \cite{sauvan2022normalization}, see Supplementary Material section 1-2.

Our design achieves a high photon-pair generation and collection efficiency  by satisfying three key conditions: (i) the presence of a mode with strong far-field directionality along the $x$-axis, enhancing out-of-plane collection; (ii) a large spatial overlap between the modal field and the pump field $\mathbf{E}_p$, which strengthens the nonlinear overlap in Eq.~(\ref{eq:G_factor}); and (iii) an eigenfrequency close to the degenerate SPDC frequency, which minimizes the detuning factor in Eq.~(\ref{eq:overlFactor}).

%The LN nano-resonator geometry investigated here satisfies these conditions. 
Simulations of the resonator, on an infinitely extended glass substrate, reveal, among the supported modes, a strongly directional QNM, labeled as \(\text{QNM}_1\). Its near-field distribution and far-field radiation pattern (calculated from the near electric field, in a closed box surrounding the resonator, with the open-source program RETOP \cite{yang2016near}) are shown in the left and right panels of Fig.~\ref{fig:concept}(b), respectively. This mode exhibits a pronounced radiation lobe directed into the substrate and aligned with the pump propagation direction, thereby satisfying condition (i).

Conditions~(ii) and~(iii) are jointly satisfied by the mode and geometry of the nanostructured resonator. First, in the near field (Fig. \ref{fig:concept}(b)), the enhanced electric field of \(\text{QNM}_1\) is polarized along the optic axis \(z\) of LN, and therefore couples with the largest element \(\chi^{(2)}_{_{\mathrm{zzz}}}\) of the second-order susceptibility tensor, maximizing the nonlinear spatial overlap. Furthermore, the chosen geometrical dimensions favor the generation of SPDC photon-pairs. Indeed, Fig.~\ref{fig:concept}(c) shows the simulated normalized photon-pair coincidence rate, obtained by considering only the QNM pair with \(m=n=1\) in Eq.~(\ref{eq:TPA}) and integrating over the signal and idler solid angles \(\Omega_\mathrm{s}\) and \(\Omega_\mathrm{i}\) within a numerical aperture (NA) of \(0.7\), over the signal angular frequency \(\omega_\mathrm{s}\) within \(\pm \,30\,\mathrm{THz}\) around the degenerate frequency \(\omega_{\mathrm{deg}}=\omega_\mathrm{p}/2\), and over all possible polarization states of the emitted photon-pairs. The coincidence rate is plotted as a function of the scaling factor~\(F_\mathrm{S}\) that is defined through
\begin{equation}
L_{\mathrm{y,z}}(F_\mathrm{S}) =
L_{\mathrm{y,z}}^{(0)}
\left(1+\frac{F_\mathrm{S}}{100}\right),
\end{equation}
where \(L_{\mathrm{y,z}}^{(0)}\) are the lateral dimensions of the geometry used in the experiment. The resonator height is kept fixed, since we aim to reproduce a realistic fabrication scenario in which the vertical dimension is primarily determined by the thickness of the thin-film platform.%Thus, \(F_\mathrm{S}=0\) corresponds to the experimental geometry, while positive and negative values indicate an increase or decrease of the lateral dimensions, respectively. 
Each count rate has been normalized by the resonator volume to account for the change induced by the factor \(F_\mathrm{S}\).

Increasing the lateral dimensions of the resonator red-shifts the QNM eigenfrequency, thereby modifying the detuning, \(\Delta \omega = (\omega_1 - \omega_{\text{deg}})/(2\,\pi)\), shown in orange on the top axis. At small detuning, a high emission rate is expected; however, detuning alone does not fully determine the emission efficiency. Variations of the resonator size also modify the spatial confinement of the mode and its interaction with the pump field, thereby affecting the spatial overlap~$|\mathcal{G}_{1,1}|$, shown by the green curve. As a result, even for comparable detuning values (e.g. \(\pm\) $7\,\mathrm{THz}$), the photon-pair count rate can vary significantly. The experimental geometry lies at the maximum of the simulated count rate, within the resolution of the explored parameter sweep. The spatial distribution of the pump electric field for this best configuration is shown in Fig.~\ref{fig:concept}(d) and shows a clear overlap with the \(\text{QNM}_1\) field, particularly in the central region of the resonator. 
Additional investigations on the choice of the resonator compared to a disk of equal size, as well as on the role of the radial cuts and the size of the inner cylinder, are reported in the Supplementary Material section 3.

In order to compare theoretical results with the experiment, a complete QNM analysis is performed. Fig. ~\ref{fig:QNMcomplete}(a) reports the real parts of the QNM eigenfrequencies together with their corresponding quality factors \(Q=  |\text{Re}(\tilde\omega_\mathrm{m})/ (2\cdot\text{Im}(\tilde\omega_\mathrm{m}))|\), where the real and imaginary parts determine the resonance frequency and the radiative leakage rate of the mode, respectively. Among these, several numerical QNMs, strongly confined within the substrate, were excluded from the final simulations (see Supplementary Material section 2 for details). Fig.~\ref{fig:QNMcomplete}(b) shows the frequency dependence of the ratio $|\mathcal{G}_{m,n}/\mathcal{S}_{m,n}(\omega_\mathrm{s})|$ for the nine QNM pairs with the largest contributions. The dominant term arises from the pair $\mathrm{QNM}_1-\mathrm{QNM}_1$. Additional significant contributions originate from QNM pairs whose eigenfrequencies are symmetrically detuned with respect to the degenerate frequency, thereby simultaneously enhancing both the signal and idler photons. Although some pairs (such as \(\text{QNM}_2\)--\(\text{QNM}_3\)) involve modes with higher Q-factors, their spatial electric-field distributions exhibit a lower overlap with the pump field, resulting in a reduced overall contribution to the SPDC process. Fig.~\ref{fig:QNMcomplete}(c) displays two typical modes (3 and 8) to illustrate the different near-field distributions and far-field radiation patterns that can arise in the resonator. Considering all retained QNM pairs and their modal overlap coefficients, the resulting SPDC distribution in wavelength and wavevector is simulated and compared with the experiment.

\section{Experiment}
\begin{figure*}
    \centering
    \includegraphics[width=\textwidth]{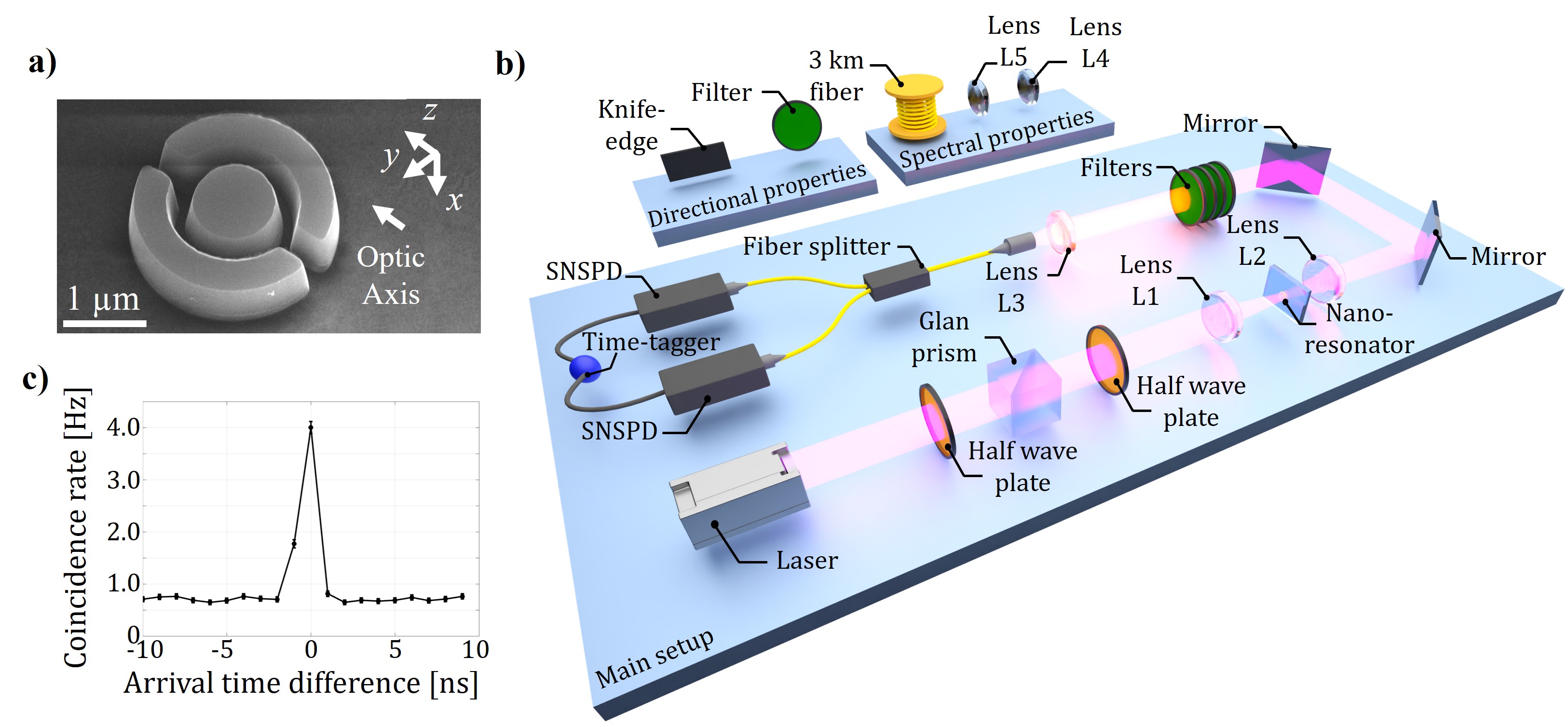}
    \caption{\textit{Experimental implementation of SPDC in the nano-resonator} (\textbf{a}) SEM picture of the nanostructured resonator (see Supplementary Material section 4). (\textbf{b}) Schematic view of the experimental setup. The continuous-wave laser power and polarization are controlled using two half-wave plates (HWPs) and a Glan–Taylor prism, and the beam is focused onto the nano-resonator by lens L1. %(focal length $f_1 = 11~\mathrm{mm}$, numerical aperture (NA) = 0.3). 
    The emitted light is collimated by lens L2, %($f_2 = 3.1~\mathrm{mm}$, NA = 0.7). 
   spectrally filtered, coupled into a single-mode fiber by lens L3, %($f_3 = 18.4~\mathrm{mm}$, NA = 0.14), 
    and photon-pair coincidences are recorded using superconducting nanowire single-photon detectors (SNSPDs) and a time tagger. Directional properties are measured both with and without spectral filtering around the degenerate wavelength. The spectrum of photon-pairs is measured by introducing a $2\times$ magnifying telescope (L4, L5) before L3 and a dispersive fiber after it. %with focal lengths of 50~mm and 100~mm, respectively. 
    (\textbf{c}) The measured distribution of the photon arrival time difference, with an incident power of 10 mW, demonstrating a coincidence peak due to the detection of photon-pairs.} %\AP{The retrieved rate is calculated by subtracting at each time delay the accidental count rate (\(\approx 0.7 \, \text{Hz} \)).}
    \label{fig:ExpSetup}
\end{figure*}

The nanostructured resonator, whose scanning-electron micrograph (SEM) is shown in Fig.~\ref{fig:ExpSetup} (a), was fabricated via focused ion beam (FIB) milling (FEI, dual-beam Helios Nanolab 650) of a commercially available \(x\)-cut 500 nm LN film on a fused silica substrate with thickness 500 \(\mu\)m~\cite{carletti2021steering}. FIB milling was preferred because it enables direct writing and precise tailoring of fine nanoscale features without requiring multiple lithography and etching steps. Patterning throughout the entire thickness was performed by using Ga$^+$ ions emitted with a current of 0.77 nA and accelerated by a voltage of 30 kV. After initial experimental characterization through second-harmonic generation, we turned to SPDC measurements. For details on the nanofabrication, SHG and preliminary SPDC characterization, see Supplementary Material Section~4.

Fig.~\ref{fig:ExpSetup}(b) presents a schematic of the experimental setup. The central panel depicts the main detection configuration, while the two upper insets illustrate the modifications employed to characterize the directional and spectral properties of the SPDC emission.

The nano-resonator was pumped from the air side by a continuous-wave laser at $725~\mathrm{nm}$ whose power was adjusted using a half-wave plate (HWP) and a Glan--Taylor prism. A second HWP adjusted the polarization along the LN optic axis. The pump beam was focused onto the resonator by lens L1 with a focal length (\(f\)) of $f_1=11~\mathrm{mm}$, producing a focal spot of approximately $5~\mu\mathrm{m}$ diameter at $\frac{1}{e^2}$ of the power, which covered the whole nanostructure. The emitted light was collimated by a second lens (L2) with  NA=0.7 and a focal length of $f_2 =3.1~\mathrm{mm}$. A cascade of long-pass spectral filters with cut-on wavelengths at 850~nm, 950~nm, 1100~nm, and 1300~nm suppressed the residual pump and isolated the SPDC emission. The filtered signal was coupled into a single-mode fiber (SMF-28) through lens L3 with focal length \(f_3 = 18.4~\mathrm{mm}\) and NA = 0.14 and divided into two paths using a 50:50 fiber beam splitter. In both paths, superconducting nanowire single-photon detectors (SNSPDs) registered photons and sent their output pulses to a time tagger. This experimental configuration, equivalent to the Hanbury Brown--Twiss detection scheme, was used to measure the pair coincidence rate. Fig.~\ref{fig:ExpSetup}(c) shows the distribution of the delay between photon detection times by the two detectors, acquired with a pump power of $10~\mathrm{mW}$. The pronounced peak around the zero delay reveals photon-pairs generated by the resonator. The total rate of pair detection per pump power is $0.45$ Hz/mW, reported by summing the counts around zero time delay and subtracting the accidental counts, measured at large time delays. The two-point coincidence peak can be attributed to imperfect temporal alignment between detectors and the finite SNSPDs jitter time. The accidental background is mainly due to uncorrelated photons from photoluminescence processes.The coincidence-to-accidental ratio (CAR) is around 4.5, demonstrating temporal correlations between the detected photon pairs. The count rate exceeds the values obtained for other nano- and microresonators \cite{marino2019spontaneous,duong2022spontaneous,saerens2023background} by more than an order of magnitude.

To characterize the angular properties of the SPDC emission, we implemented knife-edge scanning before lens L3. By recording the number of coincidences as a function of the knife edge position, we retrieved the angular  distribution of photon-pair emission. Two measurement conditions were considered: one collecting the full SPDC bandwidth (approximately $1340-1580~\mathrm{nm}$), and the second one, targeting the resonance region at the degenerate wavelength, by inserting an additional band-pass filter (BF) centered at $1450~\mathrm{nm}$ with a full width at half maximum (FWHM) of $50~\mathrm{nm}$. Both ranges are highlighted in Fig. \ref{fig:QNMcomplete}(b) with red and blue areas. 

\begin{figure*}
    \centering
    \includegraphics[width=\textwidth]{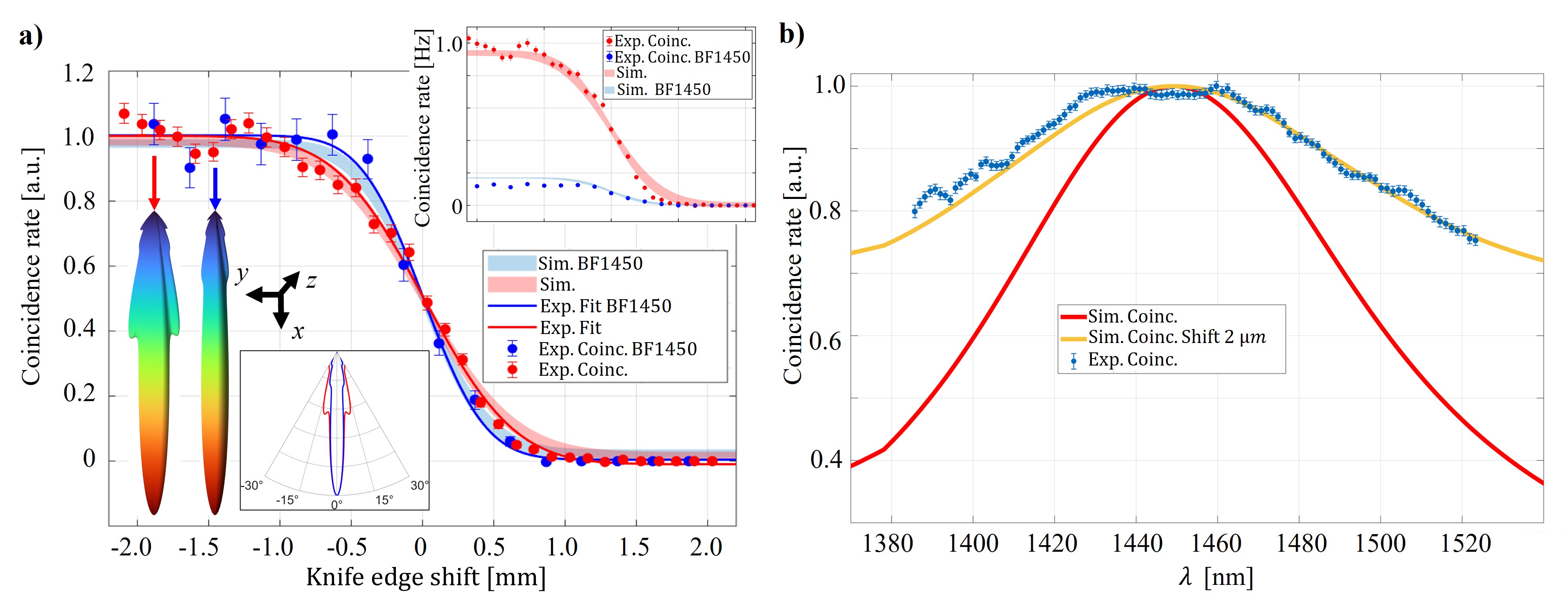}
    \caption{\textit{Angular and spectral properties of photon-pairs emission.} (\textbf{a}) Directionality of the emitted photon-pairs measured using knife-edge scans. Blue experimental points, fitted by the blue curve, are acquired with a $50$ nm band-pass filter (BF) centered at the resonance (degenerate) wavelength. Red points, fitted by the red curve, are acquired with a broader detection  bandwidth $1340-1580\,\mathrm{nm}$. The shaded regions represent the simulated knife-edge scans, with the 95\% confidence intervals of the fit. The lower insets show the far-field coincidence count rate distributions for both cases in 3D and in the plane \(x-y\), computed with an account for the limited numerical aperture (NA) of the collection optics and the single-mode fiber (SMF). The upper inset shows the absolute coincidence rates, together with the corresponding simulation predictions. The maximum coincidence rate is lower than the one reported in Fig. \ref{fig:ExpSetup}(c) due to the non-optimized collection conditions used during these measurements.
    (\textbf{b}) Experimental (blue points) and simulated (red line) spectrum of the photon-pairs. The yellow curve represents the simulated spectrum when a shift of 2 \(\mu\)m at the SMF tip is introduced between the centers of the photon-pairs emission and the Gaussian mode of the SMF.}
    \label{fig:ExpSim}
\end{figure*}

Fig.~\ref{fig:ExpSim}(a) compares the experimentally measured coincidence rates as a function of the knife-edge position with the corresponding simulated results. In the main plot, the blue points represent the normalized experimental counts obtained using the BF centered at the degenerate wavelength, while the red points correspond to measurements without the BF. The data are fitted using an error function of the form $\mathrm{erf}(x/m)$, where $x$ denotes the knife-edge displacement and $m$ characterizes the slope of the dependence. This fitting model is justified by the fact that coupling into the SMF effectively selects an approximately Gaussian spatial profile of the SPDC emission.  The resulting fits are shown as solid blue and red curves for the two measurement conditions. The obtained values of the experimental slope \(m\) are: \(m_{\text{red}} = (0.69 \pm 0.03) \text{ mm}\); \(m_{\text{blue}} = (0.49 \pm 0.06) \text{ mm}\). 

An equivalent procedure is applied to the simulated data. To account for realistic experimental conditions, Eqs.~(\ref{eq:TPA}–\ref{eq:overlFactor}) are extended to include photon collection through the lens and the SMF. The complete simulation of the SMF and its parameters are reported in Supplementary Material Section 5. The corresponding fit results are shown as shaded regions indicating the $95\%$ confidence intervals, demonstrating good agreement between experiment and simulation. The obtained parameters of the simulated curves are  \(m_{\text{red}} = (0.75 \pm 0.04) \text{ mm}\); \(m_{\text{blue}} = (0.55 \pm 0.04) \text{ mm}\). They are in good agreement with the parameters of the experimental fit. The upper inset of Fig.~\ref{fig:ExpSim}(a) shows the same curves in terms of absolute coincidence rates, together with the corresponding simulation results. A single multiplicative factor was obtained by scaling the simulated curve without BF to the corresponding experimental data and was then applied unchanged to the simulation with BF. The resulting simulated ratio of the coincidence rates between the two configurations, approximately \(1/5\), is in good agreement with the experiment. Instead, the lower insets display the corresponding 3D and polar far-field distributions of photon-pair rates, computed by accounting for the finite NA of the collection optics and the SMF used in the measurements. These insets highlight the slightly different far-field angular distributions associated with the degenerate and broadband photon-pair contributions.

Both the experimental data and the simulations show that photon-pairs generated close to the degenerate wavelength exhibit a slightly narrower knife-edge profile compared to photon-pairs collected over a broader spectral range. In bulk SPDC sources, such a behavior can be explained by phase-matching conditions, which impose a correlation between the wavelength and the angle of the emitted photons. However, in thin-film and nano-resonator SPDC sources with subwavelength thickness these constraints are largely relaxed. While in thin films this relaxation leads to no significant difference in the knife-edge scans (see Supplementary Material section 5), in nano-resonators the presence of resonant modes introduces a correlation between the spectral and spatial emission properties, resulting in two different slopes in the knife-edge scans. In particular, photon-pairs generated near the degenerate frequency are dominated by the contribution of \(\text{QNM}_1\) (see Fig. \ref{fig:QNMcomplete}(b)), whose far-field emission is highly directional.  Applying a BF centered around the eigenfrequency of \(\text{QNM}_1\) therefore selectively isolates this contribution, resulting in a sharper knife-edge transition. By contrast, when the full SPDC spectrum is collected, additional QNMs with different eigenfrequencies and spatial radiation patterns contribute to the emission, leading to a broader angular spread.

To investigate the spectral properties of the SPDC emission, we introduced a telescope formed by lenses L4 and L5 in front of lens L3 to magnify the far-field space by a factor of two, thereby improving the angular selectivity and isolating the emission associated with \(\text{QNM}_1\). In addition, we inserted a $3~\mathrm{km}$ dispersive optical fiber in the detection path to temporally stretch the two-photon wavepackets and, by mapping wavelength to time delay, retrieve the spectrum of the photon-pairs from the coincidence peak~  \cite{valencia2002entangled} (see the Supplementary Material section 6 for details).

Fig.~\ref{fig:ExpSim}(b) compares the measured and simulated SPDC spectra. The experimental data (blue points) are obtained by normalizing the spectrum measured for the resonator to that of an unstructured LN film, thereby compensating for the wavelength-dependent detection efficiency of the SNSPDs. The resulting ratio is smoothed by averaging over fourteen adjacent data points to reduce the noise and enhance the visibility of the resonance feature.
In the simulations, the $2\times$ magnification used in the experiment is taken into account. 

Compared with the simulated spectrum (red), the experimental spectrum shows a larger width and a considerable background. This discrepancy can be attributed to several factors, including a reduced quality factor of the resonator due to fabrication imperfections and, importantly, possible misalignment between the collimated SPDC emission and the SMF. Owing to the high angular selectivity of the setup, such a misalignment %In contrast, the simulations assume perfect spatial alignment between the center of SPDC emission and the center of the Gaussian mode of the SMF, resulting in the ideal selection of \(\text{QNM}_1\). 
can alter the relative modal contributions and broaden the measured resonance. This effect is illustrated by the simulation assuming a shift of the fiber mode by $2~\mu\mathrm{m}$ with respect to the SPDC center emission profile (yellow curve). In this case, the collection efficiency of \(\text{QNM}_1\) is reduced, while the contributions of other modes with different spatial profiles become more significant. The resulting spectrum is broader and has a sizable vertical offset, providing a plausible explanation for the observed discrepancy between experiment and simulation.

\section{Conclusion}
In this work, we demonstrate a highly efficient LN nanostructured resonator exhibiting a photon-pair count rate of approximately \(4.5~\mathrm{Hz}\) at an excitation power of \(10~\mathrm{mW}\), corresponding to \(0.45 ~\mathrm{Hz/mW}\). This rate exceeds by more than an order of magnitude the highest value previously reported for a single LN nanostructured resonator, namely \(0.015~\mathrm{Hz/mW}\) measured from an LN micro-cube with more than twice the lateral footprint~\cite{duong2022spontaneous}, which itself already improved by about one order of magnitude over the first nano-cylinder demonstration~\cite{marino2019spontaneous}. A broader comparison with other single-nanostructures, including normalization by volume, scattering cross section, and spectral bandwidth, is provided in Supplementary Material Section~7.
%The comparison becomes even more significant when accounting for the different resonator volumes. The nanostructured resonator has an estimated volume of \(\sim 1.75 ~\mu\mathrm{m}^3\), compared to \(\sim 46.7~\mu\mathrm{m}^3\) for the LN microcube, yielding an overall volume-normalized efficiency improvement approaching three orders of magnitude. 
%For applications where the photon-pair generation rate per unit spectral bandwidth is a relevant figure of merit, the resonator reaches a spectral brightness of approximately \(1.3~\mathrm{Hz/(W\cdot nm)}\), compared to \(\sim 0.12~\mathrm{Hz/(W\cdot nm)}\) reported for the LN microcube. 

The observed high efficiency has been understood in terms of several key aspects of the nonlinear process, including the spatial overlap between the pump field and the resonant mode, the spectral detuning from the degenerate SPDC frequency, and the directional coupling efficiency into the SMF. While the investigated geometry already satisfies these conditions to a good extent, further improvements are expected through systematic optimization of the geometrical parameters or the adoption of inverse-design strategies \cite{stich2025inverse}. Furthermore, future studies could explore more complex multi-resonant circular Bragg grating geometries for efficient non-degenerate operation~\cite{butcher2022all}.

This work extends the QNMs framework to realistic experimental scenarios and provides, to our knowledge, the first experimental test of its predictions for both directional and spectral properties of SPDC emission from a single nano-resonator. At the same time, the present study should be regarded as a first step toward a fully optimized design of nanoscale SPDC sources. Although this preliminary validation is encouraging, further experimental and theoretical investigations are required to assess the robustness and generality of the theoretical framework. In particular, further characterization of spatial, spectral, and polarization properties with different nano-resonators and metasurfaces together with further studies of SMF collection and projection will be essential to fully evaluate the capabilities and limitations of both the modelling approach and the overall nano-scale SPDC sources.

\section{Acknowledgments}
A.P. thanks Dr. Angelo Angelini for providing access to the workstation used for most of the simulations and ISSNAF for enabling the collaboration with F.M.; A.P., I.R. and M.G. acknowledge funding by NATO SPS G6026. A.Z., Y.L., M.Fi. and M.Ce. acknowledge financial support from Project NQSTI—PE\_00000023 funded by the European Union under the NextGenerationEU program - CUP H43C22000870001 Spoke 6. The work of M.P. was supported by the German Academic Scholarship Foundation. A.P. and M.Ch. acknowledge funding by European Research Council (Project 101199215 — MultiFlaQS).

\section{Author contributions}
M.P. and V.S. performed the SPDC measurements, with guidance from M.Ch and T.S. M.P. and A.Z. contributed to the design of the nanostructured resonator, with guidance from M.Ce. and M.Fi. A.P. carried out the numerical simulations with support from F.M., I.R., and M.G. A.Z. and Y.L. performed the second-harmonic-generation measurements, with assistance from M.Ce. and M.Fi. L.C. conducted preliminary simulations of the resonator. M.Fe. and A.T. fabricated the nanostructured resonator. A.P. wrote the manuscript with contributions from all authors. 

\section{Conflict of interest}
The authors declare no conflicts of interest.

\bibliography{References/references} 
\end{document}

\maketitle
\renewcommand{\thefigure}{S\arabic{figure}}

\section{Simulated properties and theoretical model}
\subsection{Geometry and nonlinear parameters}
The nanostructured resonator, made of lithium niobate (LN), was simulated with a total radius of $1.3~\mu\mathrm{m}$ and an inner-cylinder radius of 580 $\mathrm{nm}$ with a  thickness of $500~\mathrm{nm}$ and sidewalls with an angle of 80\(^\circ\). The distance between the inner-cylinder and the outer ring was 172 nm while the width of the cuts was 300 nm.\\
The nonlinear polarization was described, in the laboratory reference frame, as \cite{santiago2021photon}:
\begin{equation}
\begin{aligned}
P_\mathrm{x}^{\mathrm{NL}}(\omega_\mathrm{p}) &= 4\varepsilon_0 \Big[ d_{31}\big(E_\mathrm{x}(\omega_\mathrm{s})E_\mathrm{z}(\omega_\mathrm{i}) + E_\mathrm{z}(\omega_\mathrm{s})E_\mathrm{x}(\omega_\mathrm{i})\big) - d_{22}\big(E_\mathrm{x}(\omega_\mathrm{s})E_\mathrm{y}(\omega_\mathrm{i}) + E_\mathrm{y}(\omega_\mathrm{s})E_\mathrm{x}(\omega_\mathrm{i})\big) \Big], \\[0.5em] P_\mathrm{y}^{\mathrm{NL}}(\omega_\mathrm{p}) &= 4\varepsilon_0 \Big[
d_{22}\big(E_\mathrm{y}(\omega_\mathrm{s})E_\mathrm{y}(\omega_\mathrm{i}) - E_\mathrm{x}(\omega_\mathrm{s})E_\mathrm{x}(\omega_\mathrm{i})\big) + d_{31}\big(E_\mathrm{y}(\omega_\mathrm{s})E_\mathrm{z}(\omega_\mathrm{i}) + E_\mathrm{z}(\omega_\mathrm{s})E_\mathrm{y}(\omega_\mathrm{i})\big) \Big], \\[0.5em]
P_\mathrm{z}^{\mathrm{NL}}(\omega_\mathrm{p}) &= 4\varepsilon_0 \Big[ d_{31}\big(E_\mathrm{x}(\omega_\mathrm{s})E_\mathrm{x}(\omega_\mathrm{i})
+ E_\mathrm{y}(\omega_\mathrm{s})E_\mathrm{y}(\omega_\mathrm{i})\big) + d_{33}E_\mathrm{z}(\omega_\mathrm{s})E_\mathrm{z}(\omega_\mathrm{i}) \Big].
\end{aligned}
\label{eq:nonlinear_polarization_components}
\end{equation}
where \(E_{\mathrm{x,y,z}}(\omega)\) denote the Cartesian components of either the signal electric field, at angular frequency \(\omega_\mathrm{s}\), or the idler electric field, at angular frequency \(\omega_\mathrm{i}\), with  \(\omega_\mathrm{p}\) the pump angular frequency. The nonlinear coefficients are \(d_{22}=1.9~\mathrm{pm/V}\), \(d_{31}=-3.2~\mathrm{pm/V}\), and \(d_{33}=\frac{1}{2}\chi_{\mathrm{zzz}}^{(2)}=-19.5~\mathrm{pm/V}\) at \(1313~\mathrm{nm}\), where the contracted notation, \(d\), for \(\chi^{(2)}\) was used \cite{shoji1997absolute}. 
As a first approximation the LN was considered with two sets of refractive indices, one at the wavelength of the main resonance \((\approx 1450\) nm) : \((n_{\textrm{x}}, n_{\textrm{y}}, n_{\textrm{z}}) = (2.138, 2.211, 2.211)\) and the other at the pump wavelength (725 nm) : \((n_{\textrm{x}}, n_{\textrm{y}}, n_{\textrm{z}}) = (2.185, 2.267, 2.267)\). The refractive index of the glass substrate is set to \(n_\mathrm{g} = 1.45\), and the surrounding background is considered to be air.

\subsection{Theoretical model}
Assuming an ideal detector, the photon-pair detection rate can be written as \cite{poddubny2016generation, weissflog2024nonlinear}:

\begin{equation}
\frac{\text{d}^4 N_{\mathrm{pair}}}{\text{d}t\, \text{d}\Omega_\mathrm{i}\, \text{d}\Omega_\mathrm{s}\, \text{d}\omega_\mathrm{s}}
 = \, \mathcal{C}\left(n_\mathrm{i},\,\omega_\mathrm{p} - \omega_\mathrm{s};\,n_\mathrm{s},\,\omega_\mathrm{s}\right) \,  \left| \widetilde{T}_{\mathrm{is}}\!\left(\theta_\mathrm{i}, \varphi_\mathrm{i},\,\omega_\mathrm{p} - \omega_\mathrm{s},\,\mathbf{d}_\mathrm{i};\,\theta_\mathrm{s}, \varphi_\mathrm{s},\,\omega_\mathrm{s},\,\mathbf{d}_\mathrm{s}\right) \right|^2,
\label{eq:pair_rate}
\end{equation}
where
\begin{equation}
    \mathcal{C}\left(n_\mathrm{i},\,\omega_\mathrm{p} - \omega_\mathrm{s};\,n_\mathrm{s},\,\omega_\mathrm{s}\right) \equiv \frac{8}{\pi \mu_0^2}\, n_\mathrm{i} n_\mathrm{s}\,
\frac{(\omega_\mathrm{p} - \omega_\mathrm{s})^3 \omega_\mathrm{s}^3}{{c_\mathrm{0}}^6}
\end{equation}
and 
\begin{equation}
   \widetilde{T}_{\mathrm{is}} =  \sum_{\mathrm{m,n} = 1}^{\infty}
\xi_{\mathrm{m,n}}(\omega_\mathrm{s})\,
\tilde{E}_{\mathrm{m},\mathbf{d}_\mathrm{i}}(\theta_\mathrm{i}, \varphi_\mathrm{i})\,
\tilde{E}_{\mathrm{n},\mathbf{d}_\mathrm{s}}(\theta_\mathrm{s}, \varphi_\mathrm{s}).
\label{eq:TPA}
\end{equation}
is the complex two-photon amplitude, i.e. the probability amplitude to jointly detect the idler and signal photons in the directions given by spherical angles $ (\theta_\mathrm{i}, \varphi_\mathrm{i})$ and  $ (\theta_\mathrm{s}, \varphi_\mathrm{s})$ and with polarization vectors $\mathbf{d}_\mathrm{i}$ and $\mathbf{d}_\mathrm{s}$, respectively. The refractive indices of the medium at the idler and signal frequencies are denoted with \(n_\mathrm{i}\) and \(n_\mathrm{s}\), while \(\mu_0\) and \(c_0\) are the vacuum magnetic permeability and the speed of light, respectively. The detection rate is expressed per unit of far-field solid angles \(\Omega_\mathrm{i}\) and \(\Omega_\mathrm{s}\), as well as per unit of signal-photon angular frequency \(\omega_\mathrm{s}\). The quantity \(\tilde{E}_{\mathrm{m},\mathbf{d}_\mathrm{i}}(\theta_\mathrm{i}, \varphi_\mathrm{i})\) denotes the angular dependence of the complex far-field electric field of the \(m\)-th QNM projected onto the detection polarization \(\mathbf{d}_\mathrm{i}\). Each electric field is normalized according to the  procedure described in ~\cite{lalanne2018light} (see section 2). The summation runs over all QNM pairs (m,n), with the contribution of each pair weighted by the so-called modal-overlap coefficient,
\begin{equation}
\xi_{\mathrm{m,n}}(\omega_\mathrm{s}) \equiv \frac{\mathcal{G}_{m,n}}{\mathcal{S}_{m,n}(\omega_\mathrm{s})},
\end{equation}
where
\begin{equation}
\mathcal{G}_{\mathrm{m,n}} =
\sum_{\mathrm{\alpha,\beta,\gamma}}
\int d\mathbf r_0\,
\chi^{(2)}_{\alpha\beta\gamma}(\mathbf r_0)\,
\tilde E_{\mathrm{m,\alpha}}(\mathbf r_0)\,
\tilde E_{\mathrm{n,\beta}}(\mathbf r_0)\,
E_{\mathrm{p,\gamma}}(\mathbf r_0)
\label{eq:G_factor}
\end{equation}
and
\begin{equation}
\mathcal{S}_{\mathrm{m,n}}(\omega_\mathrm{s}) =(\omega_\mathrm{p}-\omega_\mathrm{s}-\tilde\omega_\mathrm{m})\tilde\omega_\mathrm{m}
(\omega_\mathrm{s}-\tilde\omega_\mathrm{n})\tilde\omega_\mathrm{n}.
\label{eq:overlFactor}
\end{equation}
The integral \(\mathcal{G}_{\mathrm{m,n}}\) quantifies the spatial overlap between the nonlinear susceptibility tensor \(\chi^{(2)}\), the near-field QNM eigenfields \(\tilde{\mathbf{E}}_\mathrm{m}\) and \(\tilde{\mathbf{E}}_\mathrm{n}\) along the directions \(\alpha\) and \(\beta\), and the pump field \(\mathbf{E}_\mathrm{p}\) along the direction \(\gamma\). The spectral factor \(\mathcal{S}_{\mathrm{m,n}}(\omega_\mathrm{s})\) accounts for the complex detuning between the generated photon frequencies \((\omega_\mathrm{i},\omega_\mathrm{s})\) and the QNM eigenfrequencies \((\tilde{\omega}_\mathrm{m},\tilde{\omega}_\mathrm{n})\), thereby determining the spectral and modal selectivity of SPDC. 

 To account for realistic experimental conditions, Eqs.~(\ref{eq:pair_rate}–\ref{eq:overlFactor}) are extended to include photon collection through the lens and the SMF. The SMF is modeled through a Gaussian collection function \(\phi(\mathbf{\theta}) =\exp\!\left(-w\,\tan^2(\theta)\right),\) which weights the photon-pair emission as a function of its polar angle \(\theta\). The dimensionless factor \(w\) is determined by the parameters of the SMF-28 fiber used in the experiment and the collection scheme (see section 5 for details).  The detected photon-pair count rate is calculated as
\begin{equation}
    \frac{\mathrm{d} N_{\mathrm{pair}}}{\mathrm{d}t} =  \int \mathrm{d}\omega_\mathrm{s} \int_{\text{NA}} \mathrm{d}\Omega_\mathrm{i}\, \int_{\text{NA}} \mathrm{d}\Omega_\mathrm{s} \, \phi(\theta_\mathrm{i}) \, \phi(\theta_\mathrm{s}) \,
    \mathcal{C} \, \left| \widetilde{T}_{\mathrm{is}}\!\left(\theta_\mathrm{i},\, \varphi_\mathrm{i},\,\omega_\mathrm{p} -\omega_\mathrm{s}, \, \mathbf{d}_\mathrm{i}; \,\theta_\mathrm{s},\, \varphi_\mathrm{s}, \,\omega_\mathrm{s}, \,\mathbf{d}_\mathrm{s}\right) \right|^2,
    \label{eq:final}
\end{equation}
where the variables entering the prefactor~$\mathcal{C}$ have been omitted for brevity. The integration runs only within the angles \(\theta\) allowed by the numerical aperture (NA) of the lens.

\section{QNMs study and convergence}
\subsection{QNMs normalization and computationl QNMs}
\begin{figure}[H]
    \centering
    \includegraphics[width=0.9\linewidth]{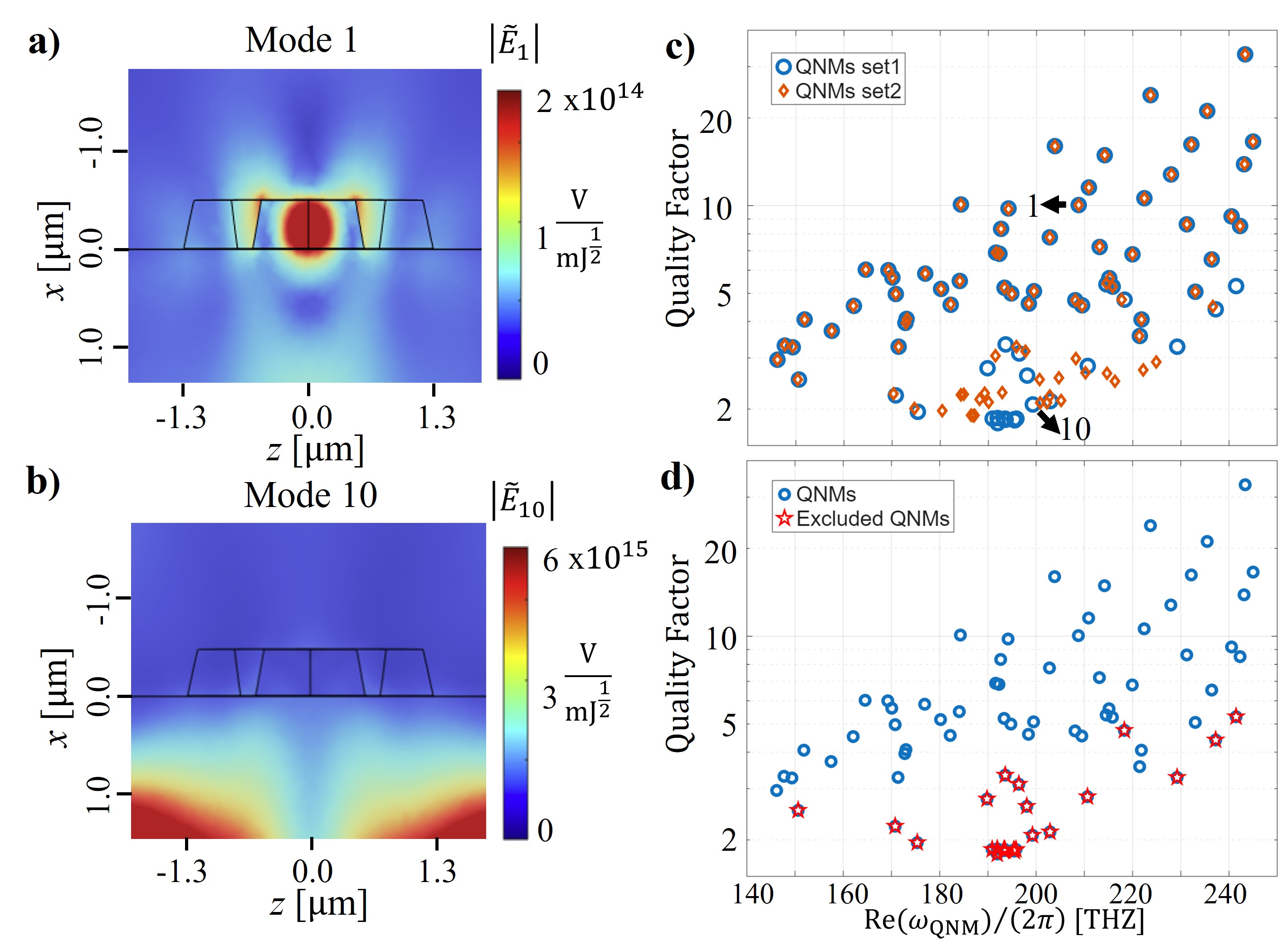}
    \caption{
    \textit{Substrate-confined QNMs and exclusion procedure.}
    (\textbf{a}) Normalized electric field amplitude \(|\tilde{E}_1|\) of \(\text{QNM}_1\), 
    showing strong confinement inside the nano-resonator. 
    (\textbf{b}) Electric-field amplitude \(|\tilde{E}_{10}|\) of a representative substrate-confined mode, with its maximum intensity located at the substrate–PML interface and negligible field inside the resonator. (\textbf{c}) Comparison of QNM eigenfrequencies and quality factors extracted using two different simulation-domain sizes. Blue circles correspond to the original domain, while orange diamonds correspond to a domain with half the lateral size and twice the vertical size. Substrate-confined modes show large frequency shifts under domain deformation. The numbers indicate the QNMs shown in panels (\textbf{a},\textbf{b}).  (\textbf{d}) QNMs included in the final expansion (blue circles) and modes excluded due to domain-size sensitivity (red stars). Only modes with normalized eigenfrequency 
    variations below \(0.1\%\) were retained.}
    \label{fig:QNMsubstrate}
\end{figure}
The QNMs were normalized according to the standard QNM normalization procedure~\cite{lalanne2018light}:
\begin{equation}
\label{eq:QNMnorm}
\langle \tilde{\mathbf{E}}_m | \tilde{\mathbf{E}}_m \rangle =
\int d\mathbf{r}\,
\Big[
\tilde{\mathbf{E}}_m(\mathbf{r}) \cdot 
\frac{\partial (\varepsilon(\omega)\,\omega)}{\partial \omega}\,
\tilde{\mathbf{E}}_m(\mathbf{r}) -\tilde{\mathbf{H}}_m(\mathbf{r}) \cdot
\frac{\partial (\mu(\omega)\,\omega)}{\partial \omega}\,
\tilde{\mathbf{H}}_m(\mathbf{r})
\Big] = 1 ,
\end{equation}
where \(\varepsilon(\omega)\) and \(\mu(\omega)\) represent the permittivity and permeability of the structure, respectively, and the derivatives are evaluated at the complex eigenfrequency \(\tilde{\omega}_m\). Here, \(\tilde{\mathbf{E}}_m(\mathbf{r})\) and \(\tilde{\mathbf{H}}_m(\mathbf{r})\) denote the electric and magnetic-field profiles of the \(m\)-th QNM and the integral is performed over the whole simulated volume including the perfectly matched layers (PML). 

Among the modes extracted from the simulation domain, several exhibited common features: they were predominantly confined within the substrate, showed negligible electric-field intensity inside the nanoresonator, and possessed relatively low quality factors. An example is shown in Fig.~\ref{fig:QNMsubstrate}(b), where the electric-field amplitude is more than one order of magnitude larger than that of \(\mathrm{QNM}_1\), shown in Fig.~\ref{fig:QNMsubstrate}(a), despite being almost entirely localized near the interface between the substrate and the perfectly matched layer (PML).\\
These modes were also found to be highly sensitive to the computational domain and PML parameters. In particular, Fig.~\ref{fig:QNMsubstrate}(c) compares the eigenfrequencies obtained using the original simulation domain (blue circles) with those obtained after halving the lateral dimension and doubling the vertical dimension (orange diamonds). While the eigenfrequencies of the resonator QNMs remain nearly unchanged, the substrate-confined modes undergo substantial shifts.\\
Such PML-sensitive modes are commonly referred to as numerical modes, or PML modes, and can contribute, together with the physical QNMs, to the completeness of the modal expansion. In the present work, however, we focus on a reduced resonant-QNM description, similar to previous nonlinear studies~\cite{gigli2020quasinormal}, of the SPDC process and therefore retain only modes whose normalized eigenfrequency variation under changes of the numerical domain is below \(0.1\%\). The excluded numerical modes are highlighted by red stars in Fig.~\ref{fig:QNMsubstrate}(d). A more complete treatment of their collective contribution to the non-resonant background represents an interesting direction for future work.

\begin{figure}[H]
    \centering
    \includegraphics[width=1\linewidth]{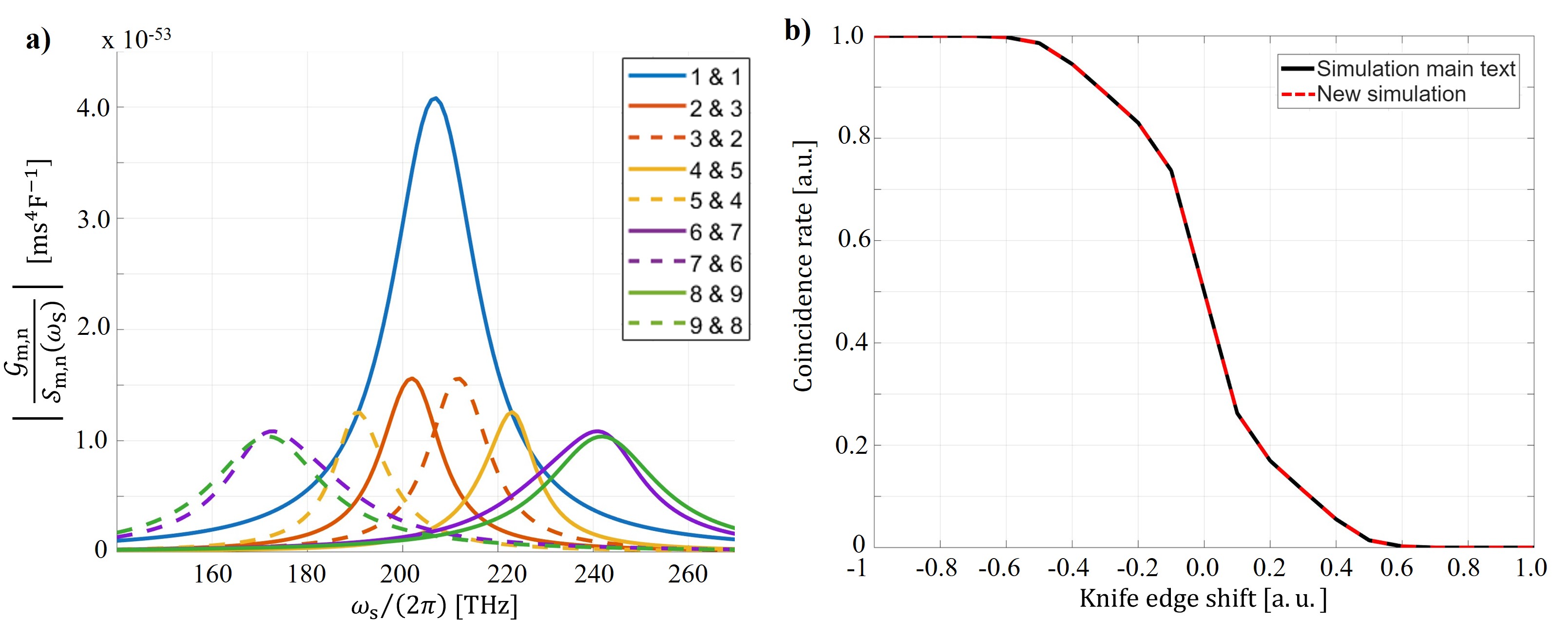}
    \caption{\textit{Robustness of the QNM calculation against variations of the computational domain and PML.} (a) Nonlinear weights \( |\mathcal{G}_{m,n}/\mathcal{S}_{m,n}(\omega_s)| \) for the dominant QNM pairs obtained with the modified numerical configuration. (b) Comparison between the normalized knife-edge response around the degenerate frequency obtained using the numerical configuration of the main text (black solid line) and the modified computational domain and PML parameters (red dashed line).}
    \label{fig:NewKnifeEdgeSimulation}
\end{figure}

To assess that the other relevant QNMs are not affected by the changes of the domain size and PML, we performed similar analysis of the main text and it is shown in Fig.~\ref{fig:NewKnifeEdgeSimulation}. Fig.~\ref{fig:NewKnifeEdgeSimulation}(a) shows the nonlinear weights of the dominant QNM pairs, calculated in the same way as in Fig.~2(b) of the main text. Both the spectral positions and the relative amplitudes of the dominant contributions are essentially unchanged, with the QNM$_1$--QNM$_1$ pair remaining the strongest contribution. Finally, Fig.~\ref{fig:NewKnifeEdgeSimulation}(b) compares the simulated knife-edge response around the degenerate frequency obtained with the original and modified numerical domains. The two curves practically overlap.
These results show that the QNMs, their relative nonlinear contributions, and the resulting observable far-field response are robust against substantial modifications of the computational domain and PML parameters.

\subsection{QNM expansion convergence}

\begin{figure}[H]
    \centering
    \includegraphics[width=1\linewidth]{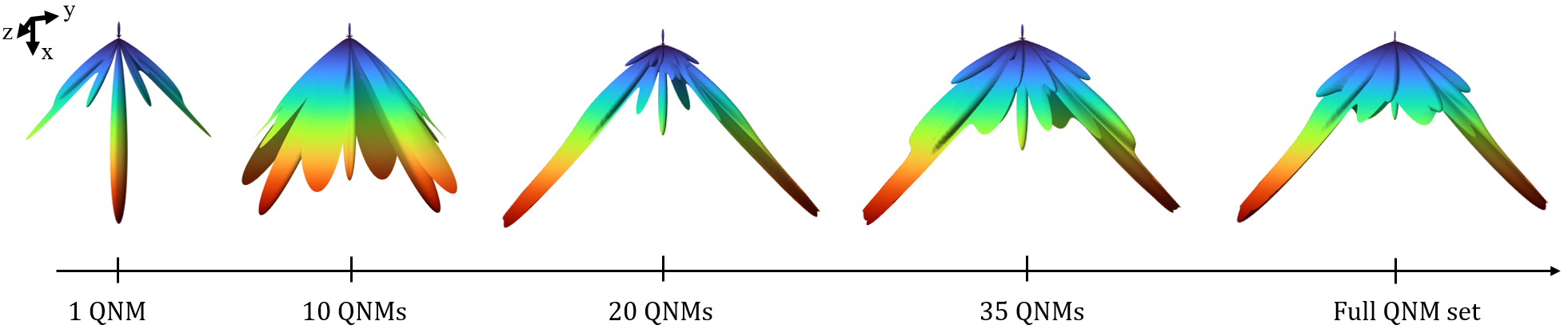}
    \caption{\textit{Convergence of the photon-pair rates expansion with QNMs.} The convergence of the expansion of Eq. (1) in the main text is studied for the degenerate and copropagating case. The QNMs are chosen taking the one with highest modal-overlap coefficient \(|\xi|\). In the leftmost panel, only the mode labeled \(\text{QNM}_1\) in the main text is considered while in the rightmost panel, all the retrieved QNMs are considered. }
    \label{fig:QNMconvergence}
\end{figure}
The convergence of the QNM expansion for linear and nonlinear optical processes has been extensively analyzed in the literature and proven in different scenarios (see Ref.~\cite{lalanne2018light} for a comprehensive review). Fig.~\ref{fig:QNMconvergence} shows the convergence of the calculated photon-pair rate as the number of included QNMs is progressively increased. The figure shows the co-propagating configuration, i.e., signal and idler photons emitted in the same direction. For clarity, the displayed photon-pair rate does not include any reduction arising from the lens system or the single-mode fiber (SMF) collection. As shown in Fig.~\ref{fig:QNMconvergence}, the overall features of the far-field photon-pair distribution are already captured when approximately 20 QNMs are included. The dominant contribution in the \(x\)-direction originates from mode \(\text{QNM}_1\) discussed in the main text, while the lateral wings arise from different radial modes. Interestingly, the strongest wing appears along the \(y\)-direction, which supports a \(z\)-polarized electric field component, whereas the wing along the \(z\)-direction is absent. This asymmetry resembles the behavior of a thin, unstructured film, where the properties of the emitted photon pairs are primarily governed by the nonlinear susceptibility tensor \(\chi^{(2)}\). In lithium niobate, whose optic axis lies along the \(z\)-direction, the dominant second-order susceptibility tensor element is \(\chi^{(2)}_{_{\mathrm{zzz}}}\). Consequently, when the structure is pumped with a \(z\)-polarized laser, photon pairs are preferentially generated with electric fields polarized along \(z\), favoring emission into directions that support this polarization component. The full calculation includes about roughly 55 QNMs, which represent the subset of valid modes remaining after removing substrate-dominated QNMs from the initial set of 80 modes.

\section{Design of the bullseye resonator}
\subsection{From a disk to the bullseye resonator}
\begin{figure}[H]
    \centering
    \includegraphics[width=0.8\linewidth]{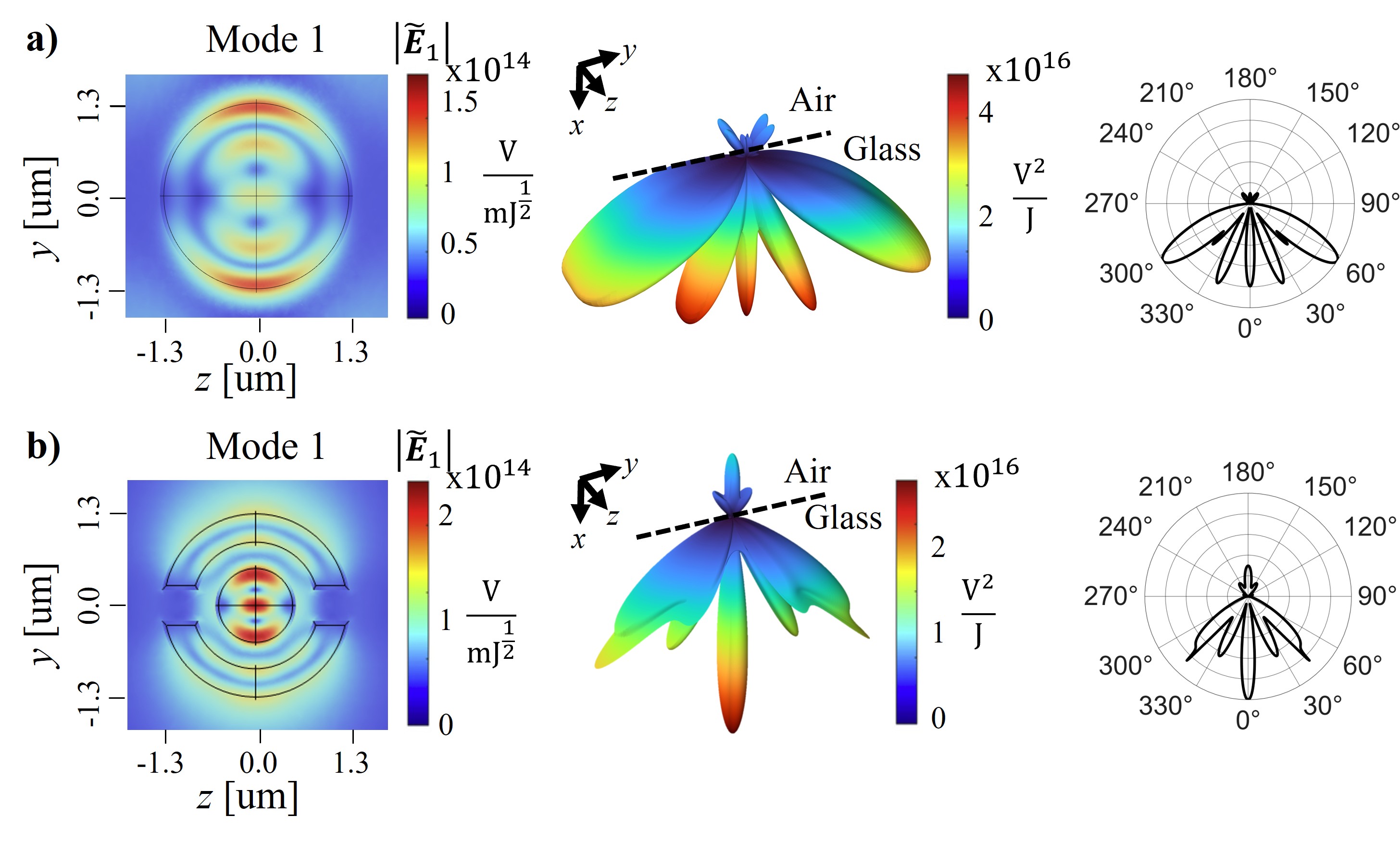}
    \caption{\textit{Comparison between a disk and the resonator studied in the main paper.} Near field distributions, 3D far-field radiation patterns, and polar emission profiles for the resonances supported by (\textbf{a}) the disk resonator and (\textbf{b}) the patterned resonator near the degenerate wavelength. }
    \label{fig:QNMdisk}
\end{figure}
The circular Bragg grating inspired resonator design was obtained starting from a disk resonator, with the goal of enhancing the emission along the propagation axis compared to other directions. This choice is motivated by the experimental difficulty of efficiently collecting photons emitted at large angles. Considering a disk resonator with the same dimensions as the experimentally used structure (radius \(\sim 1300\) nm and height \(\sim 500\) nm), the disk supports, around the degenerate wavelength (1450 nm), the resonance shown in Fig.~\ref{fig:QNMdisk} (a), exhibiting a similar spatial profile compared to the \(\text{QNM}_1\) found in the nanostructured resonator Fig. \ref{fig:QNMdisk} (b). Although this resonance exhibits an emission along the \(x\)-axis with slightly higher intensity compared to the nanostructured resonator (partially compensated by a smaller near-field intensity inside the resonator), it also presents significant emission at other angles, particularly around \(25^\circ\) and \(60^\circ\). By introducing a concentric groove in the resonator, the additional side lobes are partially suppressed with respect to the central peak, while the electric field intensity inside the resonator is simultaneously enhanced leading to a greater nonlinear interaction. The radius of the internal cut and the role of the two radial cuts are studied in the subsequent section.

\begin{figure}[H]
    \centering
    \includegraphics[width=0.6\linewidth]{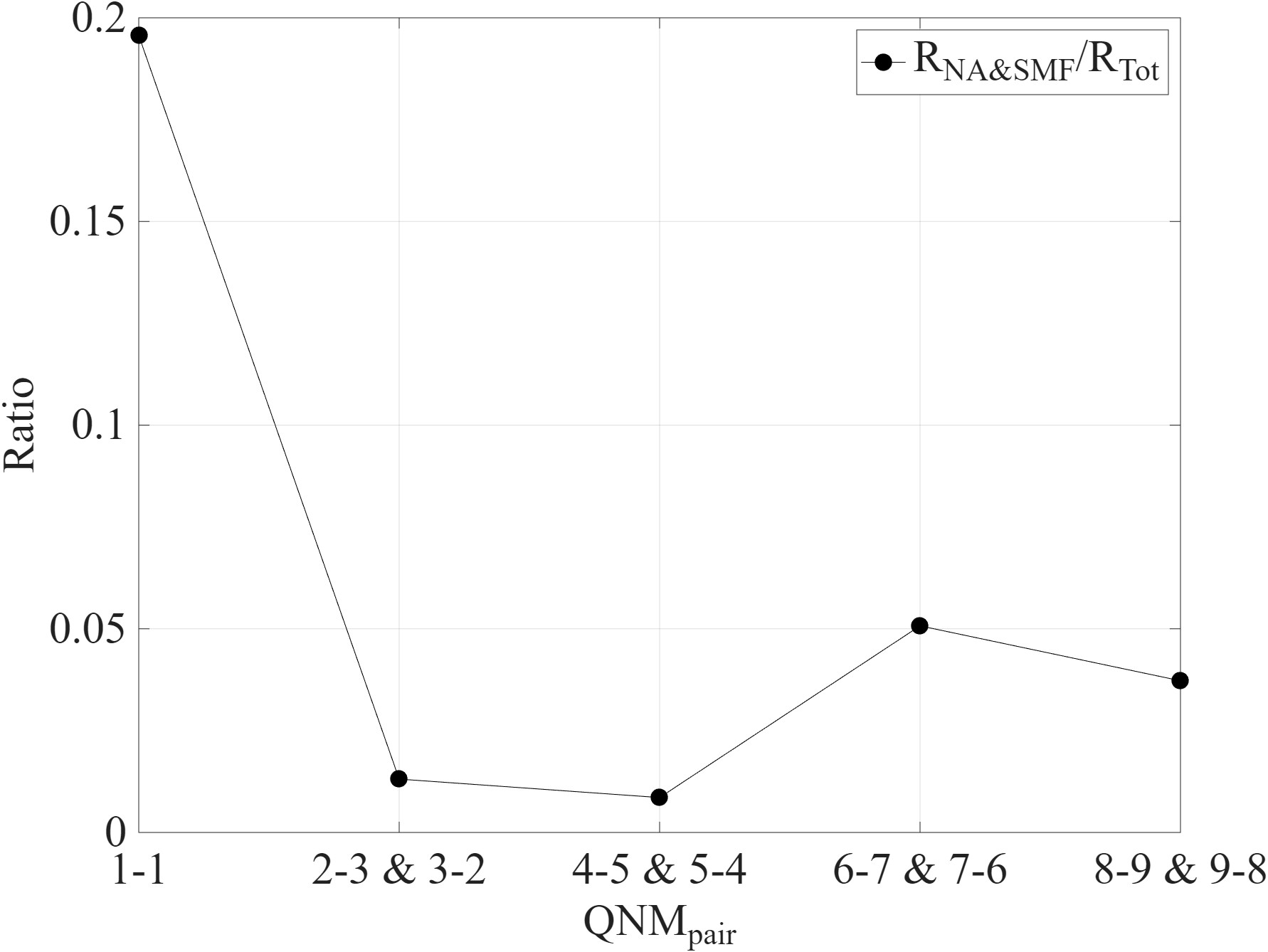}
    \caption{\textit{Collection efficiency of the dominant QNM pairs.}
    Ratio between the photon-pair rate collected within the numerical aperture and coupled to the single-mode fiber, \(R_{\mathrm{NA\&SMF}}\), and the total emitted photon-pair rate, \(R_{\mathrm{tot}}\), for the dominant QNM pairs considered in the QNM expansion. The QNM indices correspond to those reported in Fig.~2(a) of the main text.}
    \label{fig:ratio}
\end{figure}

Given the importance of this mode, we have also evaluated the directionality of the pair QNM$_1$-QNM$_1$, by evaluating the fraction of the total photon-pair emission that is collected within the numerical aperture of the optical system and coupled into the single-mode fiber. Fig.~\ref{fig:ratio} shows the ratio \(R_{\mathrm{NA\&SMF}}/R_{\mathrm{tot}}\) for the dominant QNM pairs, whose indices refer to the QNMs shown in Fig.~2(a-b) of the main text. In addition to providing the strongest contribution to the generated photon-pair rate, as shown in Fig.~2(b), the QNM$_1$--QNM$_1$ pair also exhibits the largest collection fraction, close to \(20\%\), which is substantially larger than that of most of the other relevant QNM pairs.

\subsection{Simulated emission rate as a function of the geometry parameters}
\begin{figure}[H]
    \centering
    \includegraphics[width=0.9\linewidth]{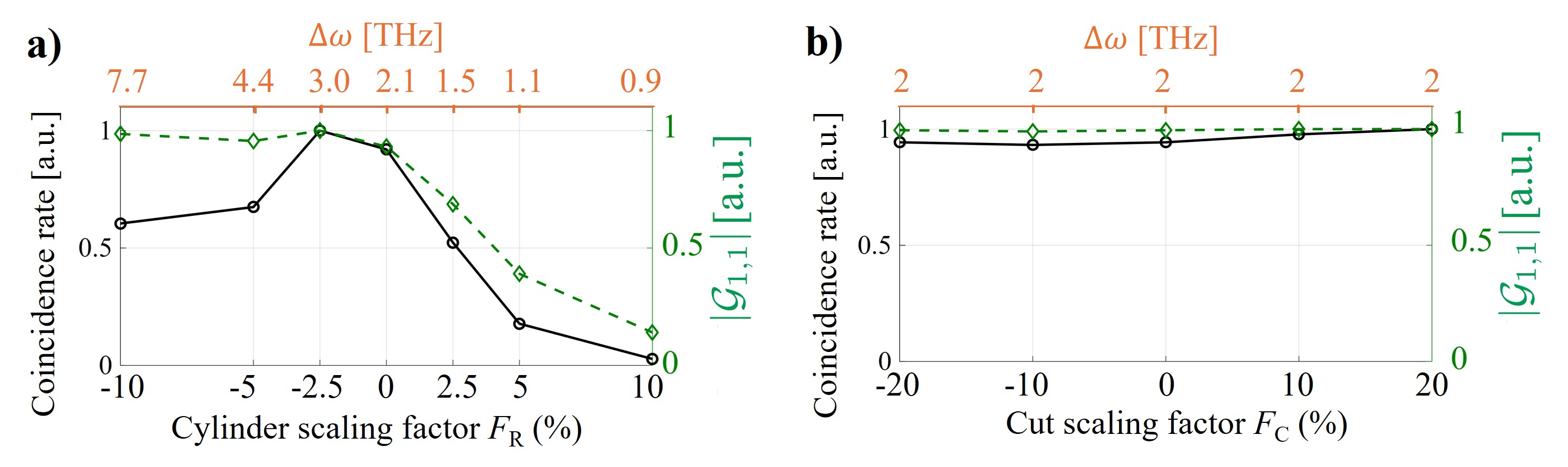}
    \caption{\textit{Additional geometrical sweeps beyond the in-plane scaling factor \(F_\mathrm{S}\).}
    (a) Normalized simulated photon-pair count rate (black), detuning from degeneracy \(\Delta\omega\) (orange, upper axis), and spatial overlap \(|\mathcal{G}_{1,1}|\) (green, right axis) as functions of the inner-cylinder scaling factor \(F_\mathrm{R}\), applied to the inner-cylinder radius in the experimentally investigated geometry. (b) Same quantities as in (a), shown as  functions of the cut scaling factor \(F_\mathrm{C}\), applied to the width of the two radial cuts. }
    \label{fig:optimization}
\end{figure}

In the main text, we focused on the in-plane lateral scaling factor \(F_\mathrm{S}\), applied to the lateral resonator dimensions. Here, we report additional geometrical studies that probe the sensitivity of the photon-pair generation rate to other structural parameters. Fig.~\ref{fig:optimization}(a) shows the effect of a scaling factor \(F_\mathrm{R}\) applied exclusively to the radius of the inner cylinder, while Fig.~\ref{fig:optimization}(b) shows the effect of a scaling factor \(F_\mathrm{C}\) applied to the width of the two radial  cuts. The central values correspond to the experimentally fabricated geometry (\(\sim 600 ~\mathrm{nm}\) for the inner cylinder radius and \(\sim 300~\mathrm{nm}\) for the cuts). In Fig.~\ref{fig:optimization}(a), the normalized photon-pair count rate varies substantially with \(F_\mathrm{R}\). In this sweep, the detuning from degeneracy remains more stable than the detuning induced by the global scaling factor \(F_\mathrm{S}\), whereas the nonlinear coupling strength \(|\mathcal{G}_{1,1}|\) changes markedly, indicating that the efficiency trend is predominantly governed by variations of the spatial and polarization overlap between the pump field and \(\text{QNM}_1\). By contrast, Fig.~\ref{fig:optimization}(b) shows that varying the cut scaling factor \(F_\mathrm{C}\) produces no appreciable change in the photon-pair count rate, in the detuning, or in \(|\mathcal{G}_{1,1}|\) within the explored range. Although the radial cuts were originally introduced to suppress radial leakage and improve vertical out-coupling, both simulations and experimental observations do not provide evidence of a significant impact on the SPDC efficiency for the presence of the radial cuts.

\section{Fabrication and first SHG and SPDC characterization}
\subsection{Nanofabrication of the resonator}
The nanostructured resonator was obtained by directly milling a commercially available x-cut LN film (thickness 500 nm) on a transparent fused-silica substrate (NanoLN, Jinan Jingzheng Electronics Co.). Focused Ion Beam machining was performed in a FEI - Dual Beam Helios Nanolab 650 system, where a Ga$^+$ ion beam was set to an emission current of 0.77 nA and an accelerating voltage of 30 kV. Before patterning, a 200 nm-thick Cr layer was deposited on top of the LN film via e-beam evaporation (PVD75, Kurt J. Lesker Company). During the ion milling process, the Cr mask prevents charging effects, reduces the implantation of Ga$^+$ ions and minimizes the presence of defects in the final structure \cite{carletti2021steering}. Then, a chemical cleaning step based on SC-1 solution (70 \% H$_2$O, 20 \% H$_2$O$_2$, 10 \% NH$_4$OH) was introduced to remove etch-induced redeposition from the nano-resonator side walls. The Cr layer was dissolved in standard etchant solution (Chrome etch 18 - micro resist technology GmbH).

\subsection{Characterization with second-harmonic generation}
\begin{figure}[H]
    \centering
    \includegraphics[width=\linewidth]{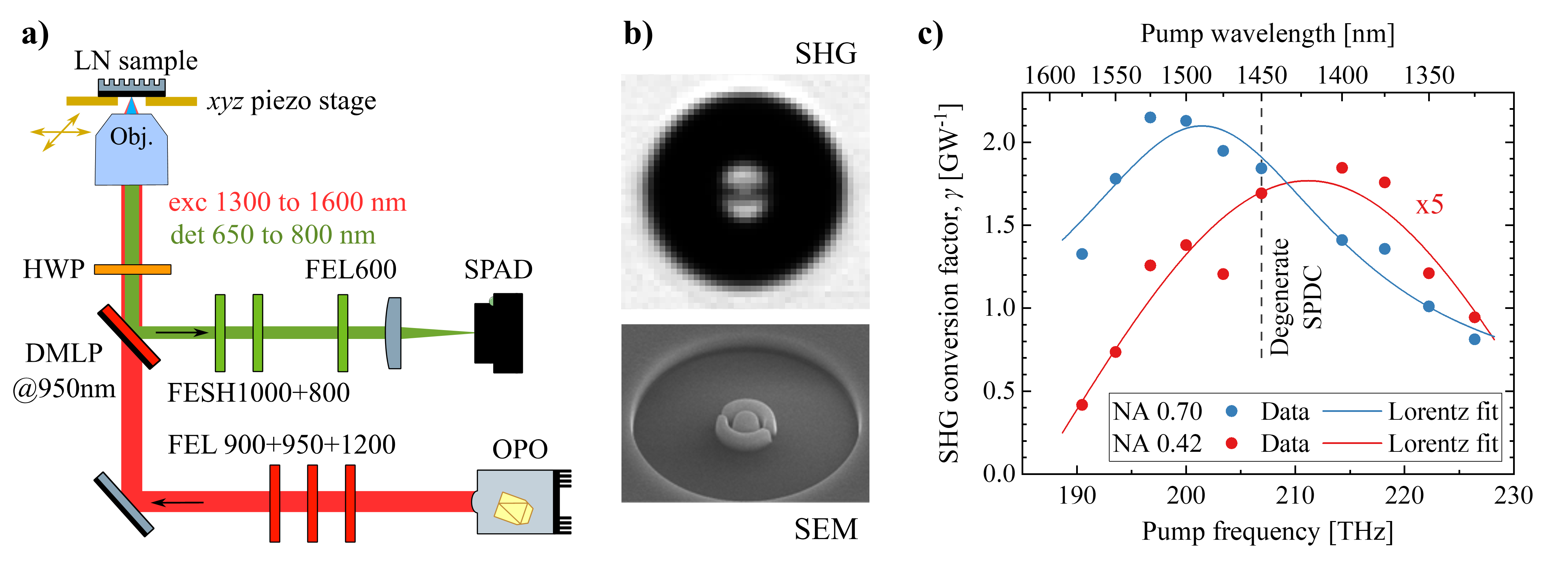}
    \caption{\textit{SHG experimental characterization of the nanostructured resonator.} (a) Simplified schematic of the experimental set-up. (b) SHG raster map (top) acquired with a 0.70 NA objective and correlated tilted-view SEM image (bottom). The circular trench dug around the resonator to bare the fused silica substrate is 10 \(\mu\)m in diameter. (c) SHG conversion factor as a function of the excitation frequency for two objectives with different NA. Dots are experimental data and solid lines Lorentzian fits.}
    \label{fig:SHG}
\end{figure}
A preliminary characterization of the nonlinear optical properties of the studied nanostructured resonator was performed via second-harmonic generation (SHG) microscopy. A simplified schematic of the experimental set-up is displayed in Fig. \ref{fig:SHG}(a). Tunable excitation over the wavelength range \(\lambda = 1325\) to \(1575\)~nm (pulse duration \(\tau \approx 150\)~fs, bandwidth \(\Delta\lambda \approx 20\)~nm, repetition rate \(f_{\mathrm{rep}} = 80\)~MHz) was provided by the signal of an optical parametric oscillator, OPO (Coherent, Chameleon Compact OPO) pumped by a Ti:Sa oscillator (Coherent, Chameleon Ultra II). Residual pump (centered \(\sim 800\)~nm) leaking into the OPO signal was suppressed through a series of long-pass filters (Thorlabs FEL900, FEL950, FEL1200). The linear excitation polarization was adjusted using an achromatic half-wave plate, HWP (Thorlabs AHWP05M-1600), to align with the \(z\)-axis of LiNbO\(_3\). Measurements were carried out in a back-scattering configuration, where both excitation and detection occurred through the same objective, on the substrate side of the sample. The collected SHG emission was separated from the excitation beam by a long-pass dichroic mirror, DMLP (Thorlabs DMLP950) and chromatically filtered to reject spurious signals (fluorescence, ambient light) through a series of edge-pass filters (Thorlabs FESH1000, FESH800, FEL600). High-sensitivity single-channel photon counting was performed by a silicon single-photon avalanche diode, SPAD (Micro Photon Devices, PD-050-CTD).

The sample was mounted onto an \(xyz\) piezoelectric stage (Physik Instrumente, P-517.3CL) to enable fine positioning of the sample and raster scanning. An exemplary SHG raster map of the resonator is shown in Fig. \ref{fig:SHG}(b), where the fine geometrical features are resolved, as highlighted by the correlated scanning electron microscopy (SEM) image presented below it. The spatial resolution is ruled by the exciting spot size scanned across the resonator and further enhanced by the nonlinear character of SHG; for this map, an objective (Nikon, CFI S Plan Fluor ELWD 60X) with a numerical aperture (NA) of 0.70 was used, resulting in a spot size \(\sim \lambda/\mathrm{NA} = 2\,\mu\mathrm{m}\) in the excitation wavelength range of interest. Similar SHG raster maps were acquired with another objective (Mitutoyo, M Plan Apo NIR 50X) with \(\mathrm{NA}=0.42\) objective, thus providing an almost even illumination of the whole structure.

The SHG efficiency of the structure was estimated by the most intense pixel of the image--namely, the most effective position of the exciting spot. After taking into account the transmission of the detection path and the efficiency of the sensor, the SHG power entering the objective can be quantified and normalized to the used excitation power (kept constant at \(P_{\mathrm{avg}}^{\omega} = 350\,\mu\mathrm{W}\)) according to the definition of the \textit{SHG conversion factor}

\begin{equation}
\gamma_{\mathrm{SHG}} \equiv \frac{P_{\mathrm{pk}}^{\mathrm{SHG}}}{\left(P_{\mathrm{pk}}^{\omega}\right)^2}
\qquad \text{with} \qquad
P_{\mathrm{pk}} = \frac{P_{\mathrm{avg}}}{\tau\,f_{\mathrm{rep}}},
\end{equation}
where \(P_{\mathrm{avg}}\) and \(P_{\mathrm{pk}}\) indicate the time-averaged and pulse-peak power, respectively. This metric captures the intrinsic nonlinear response of the structure and is independent of the excitation laser parameters, including pulse duration, repetition rate, and average power.

Conversion factor \(\gamma_{\mathrm{SHG}}(\lambda)\) is shown in Fig. \ref{fig:SHG}(c) for the resonator excited with the two objectives listed above. A resonant behavior is observed near the design wavelength of 1450~nm where degenerate SPDC is set to occur, which is underpinned by the mode \(\text{QNM}_1\). We attribute the spectral shift observed between the two resonant peaks to the excitation of different sets of modes depending on the position of the illumination spot relative to the resonator. This effect is expected to affect particularly the 0.70 NA measurement, where the spot is smaller. Indeed, as can be seen in Fig. \ref{fig:SHG}(c) , the most intense pixels occur when the beam is displaced with respect to the center of the resonator and approximately focused on the outer ring. This illumination condition may well result in the predominant excitation of resonant modes other than the targeted QNM1.

\subsection{Preliminary SPDC power-dependence measurements}
\begin{figure}[H]
    \centering
    \includegraphics[width=1\linewidth]{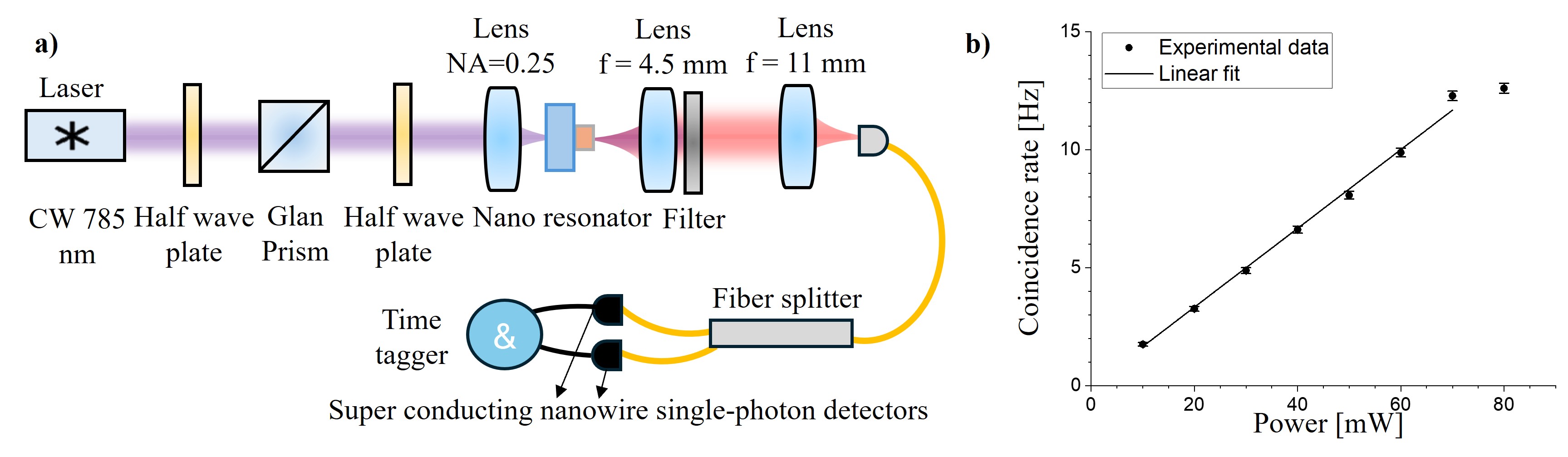}
    \caption{\textit{Preliminary SPDC power-dependence measurements.}
    (a) Experimental setup used for the preliminary characterization of SPDC from the nanostructured resonator using a CW pump at \(785~\mathrm{nm}\). (b) Measured photon-pair coincidence rate as a function of the incident pump power. The solid line represents a linear fit to the experimental data.}
    \label{fig:FirstSPDC}
\end{figure}
Fig.~\ref{fig:FirstSPDC} reports preliminary measurements of SPDC from the nanostructured resonator. The experimental configuration, shown in Fig.~\ref{fig:FirstSPDC}(a), differs from that used for the measurements presented in the main text. In particular, a CW laser at \(\lambda_{\mathrm{p}}=785~\mathrm{nm}\) was used. The pump was focused onto the resonator to a spot of approximately \(4~\mu\mathrm{m}\), while non-optimized collection optics with focal lengths of \(4.5~\mathrm{mm}\) and \(11~\mathrm{mm}\) were used to collect the emitted light and couple it into the fiber. A long-pass filter with a cut-on wavelength of \(1300~\mathrm{nm}\) was used to suppress the residual pump.

The resonator had the same geometry as that investigated in the main text, with its lateral dimensions uniformly increased by \(10\%\) (\(F_{\mathrm{S}}=10\)) to account for the longer pump wavelength. Fig.~\ref{fig:FirstSPDC}(b) shows the measured coincidence rate as a function of the incident pump power. The approximately linear dependence of the coincidence rate on pump power is consistent with the expected behavior of SPDC in the low-gain regime.

\section{Simulated spatial and spectral properties }
\subsection{Simulation of single-mode fiber}
\begin{figure}[H]
    \centering
    \includegraphics[width=0.95\linewidth]{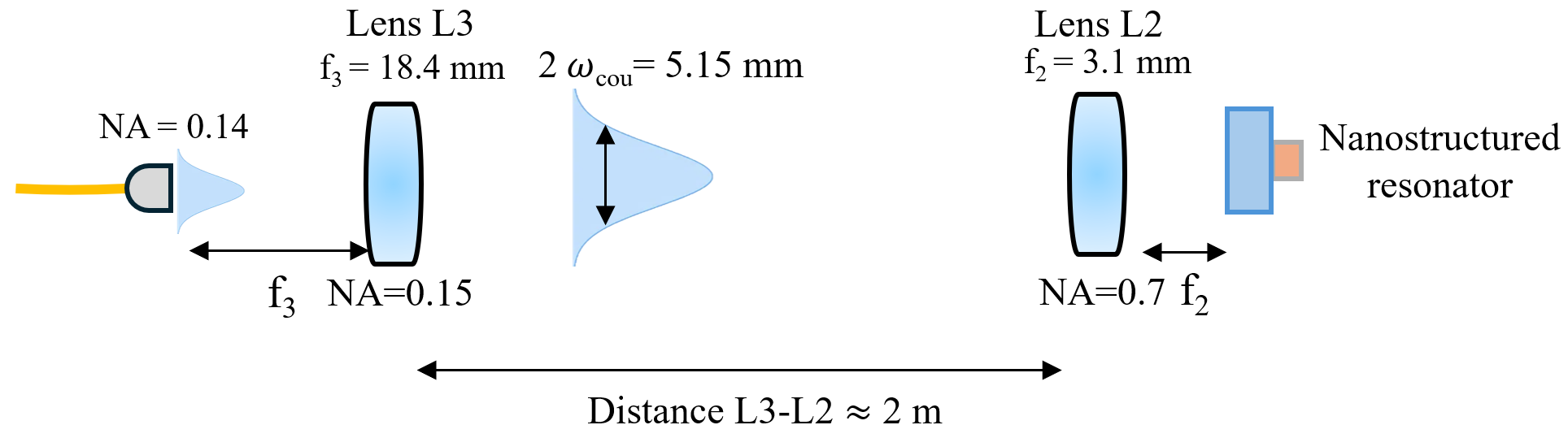}
    \caption{\textit{Experimental scheme for knife-edge measurement with single-mode fiber detection.}  Lens L3, used for fiber coupling, has a focal length of \(f_3 = \SI{18.4}{\milli\meter}\) and numerical aperture \(\mathrm{NA}_3 = 0.14\). Lens L2, employed for collimating two-photon light, has a focal length of \(f_2 = \SI{3.1}{\milli\meter}\) and numerical aperture \(\mathrm{NA}_2 = 0.7\).}
    \label{fig:scheme}
\end{figure}
\begin{figure}[H]
    \centering
    \includegraphics[width=0.95\linewidth]{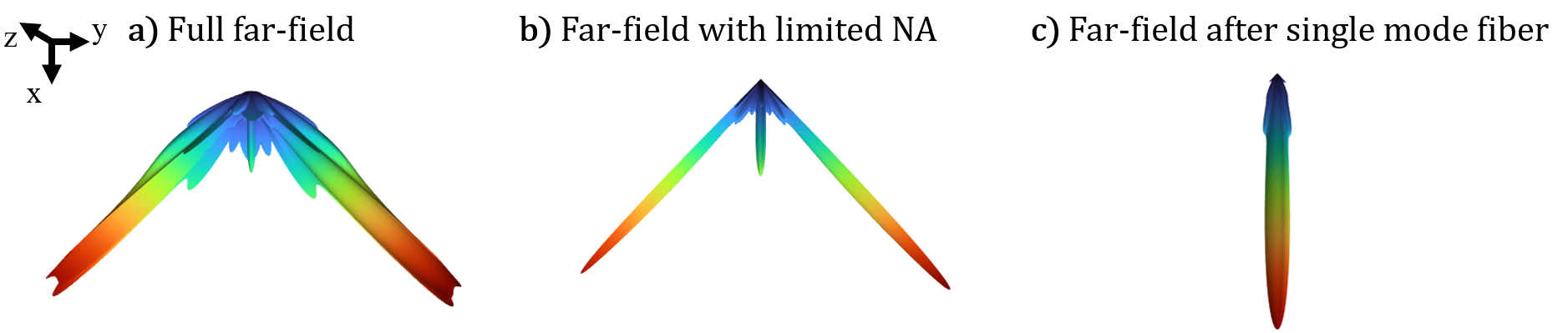}
    \caption{\textit{Photon-pair emission angular distribution for the case of degenerate co-propagating photons under different filtering 
    conditions} (\textbf{a}) full far field, (\textbf{b}) far field restricted to the NA
    of the optical system, and (\textbf{c}) far field projected on the Gaussian mode of the fiber.} 
    \label{fig:gaussianSim}
\end{figure}

%\textcolor{blue}{\it Here, you start from the fibre and consider the reverse of how the setup normally works (left to right instead of right to left). This description could be OK but would need an introduction explaining it. Instead, I am trying to keep the usual way (right to left), but then the figure should be modified, because what you call collimating lens is actually coupling lens etc.} 
To enable a meaningful comparison between the simulated and experimental data, all relevant features of the experimental setup must be included in the numerical model. The experimental configuration is summarized in Fig.~\ref{fig:scheme}. A key element is the single-mode fiber (SMF), whose spatial mode must be taken into account when evaluating the detected photon-pair rate. In particular, starting from the Gaussian mode of the fiber, we back-propagate it through the experimental setup to the nanostructured resonator. The resulting field distribution at the resonator plane is then used to evaluate the overlap integral. The Gaussian mode supported by the SMF is computed from the fiber parameters. The SMF (SMF28-Ultra) has a mode-field diameter of \(\mathrm{MFD} = \SI{6.57}{\micro\meter}\), corresponding to a Gaussian waist \(w_{0} = \mathrm{MFD}/2 = \SI{3.29}{\micro\meter}\), defined as the radius at which the field amplitude decreases to \(1/e\) (or intensity to \(1/e^2\)) of its peak value. The fiber numerical aperture is \(\mathrm{NA}_\mathrm{fiber} = 0.14\), and the operating wavelength is \(\lambda \approx \SI{1440}{\nano\meter}\). The Rayleigh range of the fundamental fiber mode is
\begin{equation}
    x_R = \frac{\pi w_0^2}{\lambda} \approx \SI{23.5}{\micro\meter},
\end{equation}
which is much shorter than the propagation distances involved in the experiment (\(f_3 = 18.4 \text{ mm} \)). Therefore, the fiber NA can be used directly to estimate the Gaussian waist of the beam on lens L3:
\begin{equation}
    w_{\mathrm{cou}} %= f_3 \tan\theta 
    \approx f_3 \,\mathrm{NA}_{\mathrm{fiber}}
    \approx \SI{2.58}{\milli\meter}.
\end{equation}
The corresponding Rayleigh range before L3 is
\begin{equation}
    x_R = \frac{\pi w_{\mathrm{cou}}^2}{\lambda} \approx \SI{14}{\meter},
\end{equation}
confirming that the field coupled into the fiber is effectively collimated along the experimental path. 
The corresponding intensity distribution at the plane of lens \(L_2\) is described by the Gaussian function \(\phi(\boldsymbol{\rho})=\exp\!\left(-\frac{2\,|\boldsymbol{\rho}|^2}{w_{\mathrm{cou}}^2}\right)\), where \(|\rho|\) is the radial distance.
To overlap this mode with the far-field emission of the resonator, the transverse coordinate is expressed in terms of the emission angle \(\theta\) through the geometrical relation \(|\boldsymbol{\rho}| = f_2 \tan\theta\) valid for a source located at the focal point of lens \(L_2\). The mode collected by the SMF can therefore be rewritten in angular coordinates as
\begin{equation}
\phi(\theta)=\exp\!\left(-\frac{2\,f_2^2}{w_{\mathrm{cou}}^2}\tan^2\theta\right).
\end{equation}
In the main text the parameter \(w\) is thus defined as \(w = \frac{2\,f_2^2}{w_{\mathrm{cou}}^2} \). Fig. ~\ref{fig:gaussianSim} compares the photon-pair emission pattern for the degenerate, co-propagating configuration under three different conditions: (\textbf{a}) the full angular intensity distribution without any filtering, (\textbf{b}) the angular intensity distribution restricted to the NA of the optical system,  and (\textbf{c}) the angular intensity distribution after the projection on the Gaussian mode of the SMF. As shown, the SMF mode imposes a strong spatial filtering effect on the detected photon pairs, making its inclusion essential for accurate comparison between theory and experiment.

\subsection{Simulated angular properties: calibration with the unstructured film}
\begin{figure}[H]
    \centering
    \includegraphics[width=0.5\linewidth]{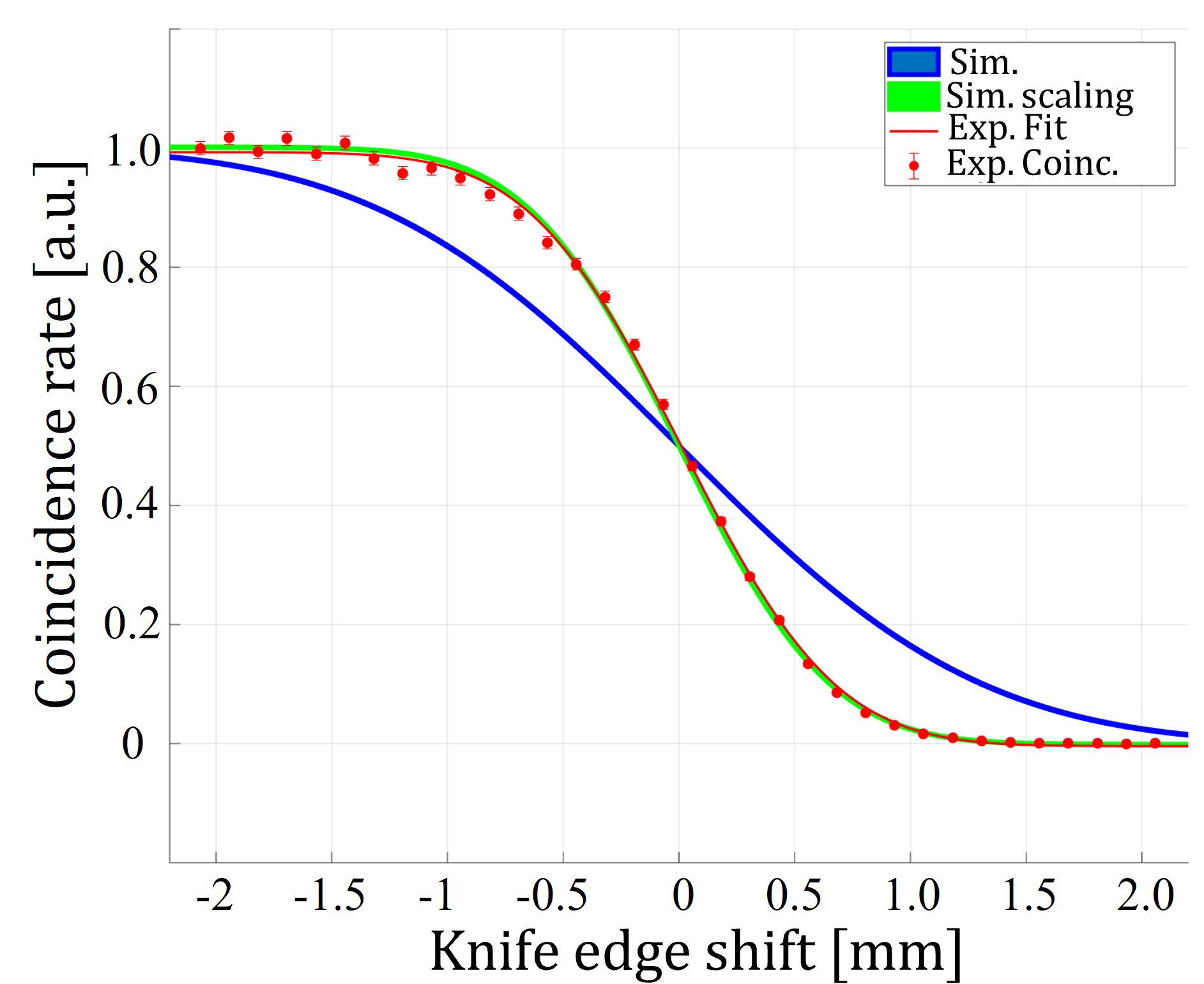}
    \caption{\textit{Comparison between the simulated and experimental knife-edge measurements for a thin film.} Comparison between the experimental knife-edge trace from the unstructured film and the simulated uniform distribution including NA limitation and Gaussian fiber overlap with and without the introduction of the scaling factor.}
    \label{fig:ExpSimFilm}
\end{figure}

The initial calibration of the knife-edge model was performed by considering the isotropic emission from an unstructured LN film having the same thickness as the resonator, for which the far-field angular distribution is analytically known. The corresponding knife-edge response was simulated by explicitly accounting for both the numerical aperture of the collection optics and the spatial filtering imposed by the SMF. The simulated isotropic knife-edge curve was then compared with the experimental knife-edge measurement obtained from a \(500~\mathrm{nm}\)-thick LN film. The results are reported in Fig.~\ref{fig:ExpSimFilm}, where the simulated curve is shown in blue and the experimental data points in red together with their fit. A clear discrepancy is observed, with the simulated knife-edge response appearing broader than the experimental one. We attribute this mismatch to an effective beam narrowing in the experimental setup, since the knife-edge measurement was performed immediately before collection lens L3, located several meters from lens L2. Even slight residual focusing or beam compression over this propagation distance can therefore produce a significant reduction of the beam diameter at the knife-edge plane. To account for this effect, a scaling factor was introduced as a free parameter to rescale the simulated knife-edge position and match the experimental thin-film measurement. The resulting corrected simulation is shown in green in Fig.~\ref{fig:ExpSimFilm}. The same scaling factor is also applied to the knife-edge curves reported in the main text for the nanostructured resonator. Since this parameter is independently retrieved from the thin-film calibration, it does not affect the physical interpretation of the resonator results.

\subsection{Simulation of the spatial properties and knife-edge scans for thin films}
\begin{figure}[H]
    \centering
    \includegraphics[width=0.95\linewidth]{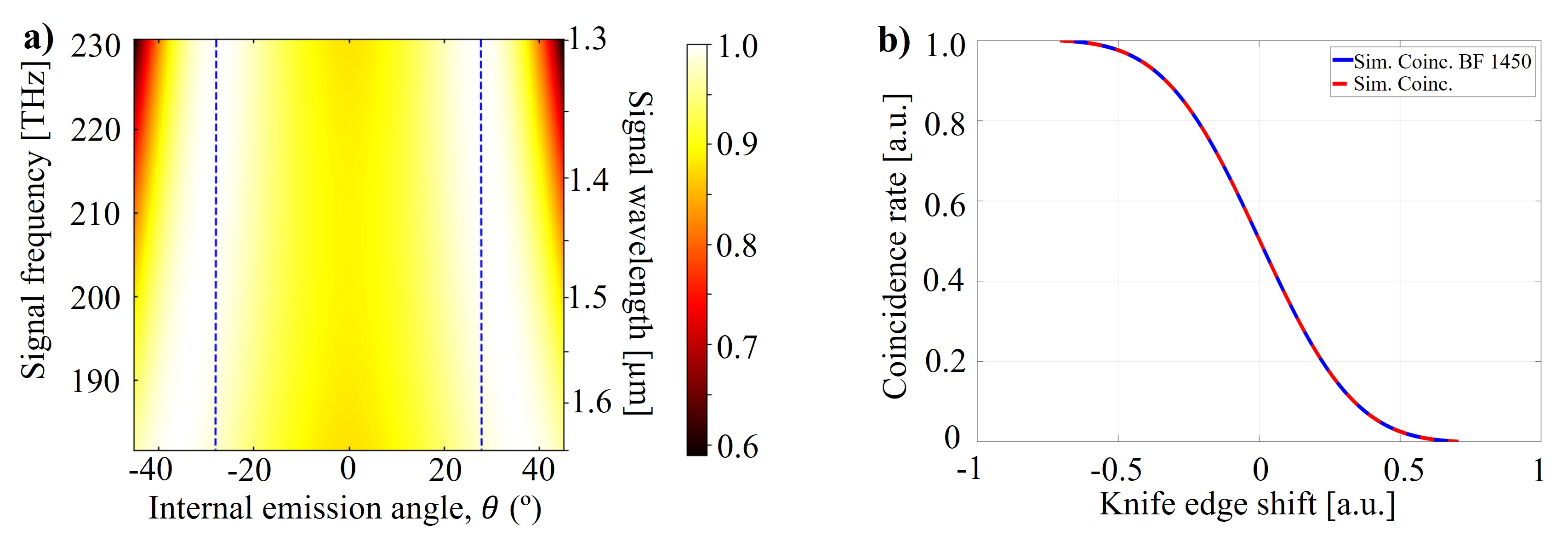}
    \caption{\textit{Thin-film simulated coincidence rate and knife-edge scans.} (\textbf{a}) Simulated signal spectrum in a 500 nm thick lithium-niobate thin film as a function of signal frequency and internal emission angle. The dashed blue lines denote the total internal reflection angle at the amorphous-silica interface. (\textbf{b}) Corresponding simulated knife-edge scans for photon pairs collected around the degenerate wavelength and over the full spectral range, showing overlapping profiles. The knife edge curves are evaluated by considering the SMF and the reduced NA of the lens of the setup.}
    \label{fig:ThinfilmKnife}
\end{figure}

In the main text, the simulated and experimental results show a different behaviour when considering photon pairs collected within 50 nm around the degenerate wavelength 1450 nm compared to the full spectral range (\(\approx 1340-1580\) nm). 
This difference originates from the modal properties of the resonator and distinguishes it from the case of a thin unstructured film. In Fig.~\ref{fig:ThinfilmKnife}(a) we show the simulated signal spectrum retrieved from Ref.~\cite{sorensen2025simple}, for a lithium-niobate thin film with a thickness of 500 nm, plotted as a function of the signal frequency (or wavelength) and the internal emission angle. As visible, the emission profile shows almost no dependence on the wavelength due to the complete relaxation of the phase-matching condition in the thin film. As a consequence, the knife edge scans corresponding to the two spectral collection configurations completely overlap, as shown in Fig.~\ref{fig:ThinfilmKnife}(b).

%\subsection{Experimental absolute counts for knife-edge measurements}
%\begin{figure}[H]
%    \centering
%    \includegraphics[width=0.75\linewidth]{Images/Supplementary/Knife_Edge_No_Normalization.jpg}
%    \caption{\textit{Absolute counts for the knife-edge measurements.} The red points show the photon-pair count rate acquired without any additional spectral filter, while the blue points show the count rate obtained with a band-pass filter centered around \(1450~\mathrm{nm}\). The red and blue solid lines represent the corresponding fits obtained using an error function.}
%    \label{fig:KnifeEdgeCounts}
%\end{figure}
%In Fig.~\ref{fig:KnifeEdgeCounts}, the absolute count rates obtained with a pump power of \(10~\mathrm{mW}\) are reported without any normalization. The count rate around degeneracy constitutes approximately one fifth of the total broadband count rate. The maximum broadband count rate does not reach the maximum value reported in the main text (\(4.5~\mathrm{Hz}\)) because the collection conditions were not fully optimized during the acquisition of the reported measurements.

\section{Experimental spectral properties: calibration curve}

To measure the spectrum of photon pairs, we use two-photon spectroscopy \cite{valencia2002entangled,santiago2021entangled} based on propagation through a dispersive medium. In this approach, photon pairs traverse a long dispersive fiber before being detected. Group-velocity dispersion (GVD) spreads two-photon wavepackets in time while preserving their spectral content. For a monotonic GVD with fixed sign over the wavelength range of interest, a one-to-one correspondence is established between the arrival-time difference of the two photons and their wavelengths. As a result, the coincidence histogram directly maps temporal delays onto the SPDC spectrum, enabling spectral reconstruction without the use of single-photon spectrometers. The exact value of the GVD depends on the specific fiber and can vary due to fabrication tolerances; furthermore, the effective fiber length is known only within a limited accuracy. For this reason the fiber-based spectrometer is calibrated experimentally. This is achieved by recording coincidence histograms using different filters with a known cut-on wavelength and matching the edges of the known transmission spectra of the filters with the edges of acquired coincidence histograms.

\begin{figure}[H]
    \centering
    \includegraphics[width=0.95\linewidth]{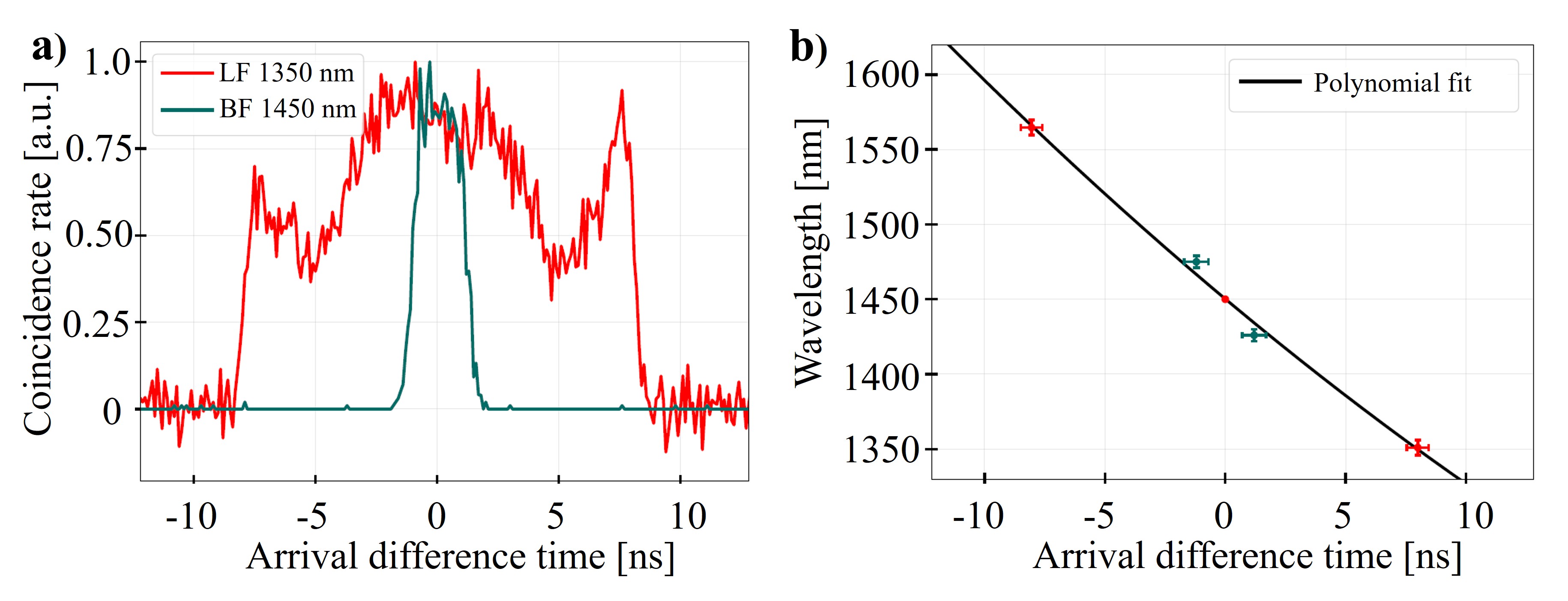}
    \caption{\textit{Calibration of spectral measurements.} (\textbf{a}) Coincidence histogram recorded using a long-pass filter with a cut-on wavelength of 1350 nm (red) and a band-pass filter centred at 1450 nm (green). (\textbf{b}) Calibration curve (black) for mapping the arrival time difference onto the wavelength.}
    \label{fig:Spectral}
\end{figure}

Specifically, coincidence histograms are acquired using a long-pass filter with a cut-on wavelength of 1350~nm and a band-pass filter centered at 1450~nm with a full width at half maximum (FWHM) of 50~nm, as shown in Fig.~\ref{fig:Spectral}(a). Owing to the positive GVD of the fiber, the right edge of each histogram (positive delays) corresponds to the filter cut-on wavelength. Using the energy conservation in SPDC and the known pump wavelength, the conjugate wavelength of the photon pair is inferred and associated with the opposite edge of the histogram.

The resulting wavelength -- time delay pairs are used to construct a calibration curve, shown in Fig.~\ref{fig:Spectral}(b), by fitting with a polynomial function. The uncertainties in the calibration points account for both the finite spectral width of the filter edges and the timing jitter of the detectors. This calibration enables a direct conversion of arrival-time differences into wavelength, allowing reconstruction of the SPDC spectrum from the measured coincidence histograms.

\section{Comparison with other single-nanostructure SPDC sources}
\begin{table*}[h]
\centering
\caption{\textit{Comparison of experimentally demonstrated SPDC sources based on isolated nanostructures.}
\(R_{\mathrm{det}}/P_{\mathrm{pump}}\) denotes the detected photon-pair rate normalized to the incident pump power, \(V\) is the physical volume of the nanostructure, \(B\) is the spectral brightness, and \(\sigma_{\mathrm{sca}}\) is the scattering cross section at the pump wavelength. A dash indicates that the corresponding quantity is not available or was not evaluated. For Ref.~\cite{duong2022spontaneous}, the highest-rate microcube used for the comparison in the main text is reported separately from microcubes 2 and 3, for which the scattering cross section could be numerically evaluated.}
\label{tab:resonator_comparison}
\resizebox{\textwidth}{!}{
\begin{tabular}{lcccccc}
\hline
Structure &
\(R_{\mathrm{det}}/P_{\mathrm{pump}}\) &
\(V\) &
\(\frac{R_{\mathrm{det}}}{P_{\mathrm{pump}}V}\) &
\(B\) &
\(\sigma_{\mathrm{sca}}\) &
\(\frac{R_{\mathrm{det}}}{P_{\mathrm{pump}}\sigma_{\mathrm{sca}}}\)
\\
 &
\([\mathrm{Hz/mW}]\) &
\([\mu\mathrm{m}^{3}]\) &
\([\mathrm{Hz/(mW\cdot\mu m^{3})}]\) &
\([\mathrm{Hz/(W\cdot nm)}]\) &
\([\mu\mathrm{m}^{2}]\) &
\([\mathrm{Hz/(mW\cdot\mu m^{2})}]\)
\\
\hline

LN resonator (this work)
& \(0.45\)
& \(1.8\)
& \(2.6\times10^{-1}\)
& \(1.3\)
& \(14\)
& \(3.3\times10^{-2}\)
\\

LN microcube, highest-rate device~\cite{duong2022spontaneous}
& \(0.015\)
& \(47\)
& \(3.2\times10^{-4}\)
& \(0.12\)
& --
& --
\\

LN microcube 2~\cite{duong2022spontaneous}
& \(0.0014\)
& \(16\)
& \(9.0\times10^{-5}\)
& \(0.011\)
& \(26\)
& \(5.5\times10^{-5}\)
\\

LN microcube 3~\cite{duong2022spontaneous}
& \(0.0040\)
& \(22\)
& \(18\times10^{-5}\)
& \(0.032\)
& \(19\)
& \(2.1\times10^{-4}\)
\\

AlGaAs nanocylinder~\cite{marino2019spontaneous}
& \(0.005\)
& \(0.06\)
& \(8.3\times10^{-2}\)
& --
& \(0.93\)
& \(5.4\times10^{-3}\)
\\

GaAs nanowire~\cite{saerens2023background}
& \(0.002\)
& \(0.8\)
& \(2.5\times10^{-3}\)
& \(0.016\)
& \(1.5\)
& \(1.3\times10^{-3}\)
\\

\hline
\end{tabular}
}
\end{table*}

A direct quantitative comparison between nanoscale SPDC sources is not straightforward, since the detected photon-pair rate depends on many properties of the resonator and the type of detection. For this reason, no single normalization can provide a complete measure of the performance of a nonlinear nanoresonator.
Table~\ref{tab:resonator_comparison} summarizes the experimentally demonstrated single-nanostructure SPDC sources using different figures of merit. Compared to the main text, we further introduce an alternative normalization based on the optical scattering cross section at the pump wavelength, as a measure of the interaction of the incident pump field with the nanostructure.

In absolute detected photon-pair rate per pump power, the present resonator reaches \(0.45~\mathrm{Hz/mW}\), corresponding to an enhancement of approximately \(30\) compared with the highest-rate LN microcube~\cite{duong2022spontaneous}, \(90\) compared with the AlGaAs nanocylinder~\cite{marino2019spontaneous}, and \(225\) compared with the GaAs nanowire~\cite{saerens2023background}. Strong enhancement is observed for the other figures of merit. The rate normalized by the scattering cross section was not evaluated for the highest-rate microcube because of the high computational requirements associated with simulating its larger geometry. For the AlGaAs nanocylinder, the experimentally measured spectrum was not available from the reported data, preventing a reliable evaluation of the corresponding spectral brightness.

\bibliography{References/references}